# Computational predictions of nutrient precipitation for intensified cell culture media via amino acid solution thermodynamics

Jayanth Venkatarama Reddy[1], Nelson Ndahiro[1], Lateef Aliyu[1], Ashwin Dravid[1], Tianxin Xang[1], Jinke Wu[1], Michael Betenbaugh[1*], and Marc Donohue[1*]

1) Department of Chemical and Biomolecular Engineering, Johns Hopkins University, Baltimore, MD, 21218, USA

*Correspondence:
Marc Donohue: mdd@jhu.edu
Michael Betenbaugh: beten@jhu.edu, mjbeten@gmail.com

# Abstract

The majority of therapeutic monoclonal antibodies (mAbs) on the market are produced using Chinese Hamster Ovary (CHO) cells cultured at scale in chemically defined cell culture medium. Because of the high costs associated with mammalian cell cultures, it is desired to run these cultures with high cell densities to produce high product titers. These bioprocesses require high concentrations of nutrients in the basal media and periodically adding concentrated feed media to sustain cell growth and therapeutic protein productivity. For these bioprocesses, the cell culture media can contain 50 to 100 compounds. Unfortunately, the desired or optimal nutrient concentrations of the feed media are often solubility limited due to precipitation of chemical complexes that form in the solution. Experimentally screening the various cell culture media configurations can involve formulating the 50 to 100 compounds at different concentrations; this can be expensive and laborious. This article lays the foundation for utilizing computational tools to understand precipitation of nutrients in the context of cell culture media by studying the interactions between two amino acids in water via thermodynamic models. Activity coefficient data for binary systems of one amino acid in water and amino acid solubility data in ternary systems of two amino acids in water have been used simultaneously to determine a single set of UNIFAC interaction parameters to predict the thermodynamic behavior of the multi-component systems found in mammalian cell culture media. The data collected in this study is, to our knowledge, the largest set of ternary system amino acid solubility data reported to date. The predictions of solubilities of amino acids have been verified against experimentally measured amino acid solubilities in ternary and quaternary amino acid solutions. Thus, we demonstrate the utility of our model as a digital twin to identify optimal cell culture media compositions by replacing empirical approaches for nutrient precipitation with computational predictions based on thermodynamics of individual media components in complex mixtures.

# 1 Introduction

Bio-therapeutics have been used successfully to treat various diseases including several types of cancer, autoimmune disorders, infectious diseases such as HIV, COVID-19, etc. The success of using bio-therapeutics to treat these conditions can be attributed to their biocompatibility and their ability to target specific antigens [1-3]. A large number of bio-therapeutics have been approved by the US FDA and the European Medicines Agency with monoclonal antibodies (mAbs) being the majority of those approved. In 2017 the mAb market was $98 billion and this grew to over $340 billion by 2021 [4] and over $500 billion in 2024 [5]. However, bio-therapeutics still are expensive with treatments often costing $10,000 or more and in some cases exceeding $500,000 per treatment [6]. To meet the growing global demand for these therapies at affordable costs, there is an immediate need for process intensification. Process intensification in biomanufacturing involves increasing volumetric productivity by optimizing cell culture media, bioreactor operations, and cell lines [7]. Biologics typically are produced in living cells with Chinese Hamster Ovary (CHO) cells representing the primary production platform [8]. These cells often are grown in fed-batch mode in which cells are initially cultivated in basal media and periodically pulsed with concentrated feed media at regular intervals to prevent depletion of essential nutrients. Cell culture media often contains 50 to 100 compounds including salts, amino acids, sugars, lipids, and in some cases, proteins [9]. Development of cell culture media for intensified processes poses several challenges. Principal among these is the desire to dissolve larger concentrations of nutrients into the basal media and feed formulations; however, this can result in precipitation of amino acids and other components [10]. Precipitation of amino acids in feed media has been shown in supplementary Figure S1. These solubility and stability issues can detract from the ability to maintain a consistent bio-manufacturing process while also limiting the availability of key nutrients in the required quantities, leading to suboptimal bioprocess performance [9]. Therefore, precipitation is problematic from both product yield and regulatory perspectives.

A major goal of media design for intensified bioprocessing would be to identify and alter the levels of those components that are problematic in terms of solubility and precipitation. However, identifying which components will precipitate is the first step in changing composition(s). Unfortunately, this is challenging because the precipitate can contain multiple compounds and is further complicated by the scarcity of studies that quantify the solubilities of these compounds in water and in complex multi-component media formulations. Furthermore, until the development and publication of AMBIC 1.0 media and feed composition [11] nearly all cell culture media formulations were proprietary. Recently, a study quantifying the precipitate of AMBIC 1.0 feed media revealed the presence of a precipitate containing 8% to 13% bound water, 80% to 85% volatile content (assumed to be organic), and less than 13% nonvolatile content (assumed to be inorganic). The organic content was further analyzed and found to include tyrosine (77% Wt%), phenylalanine (4% Wt%), and 8 other amino acids in trace amounts. The inorganic content mostly contained metal sulphates of iron, manganese, sodium, and zinc [12]. The presence of these crucial nutrients in the precipitate affects their bioavailability and this can have a significant negative effect on the performance of the cells in culture and ultimately the critical quality attributes (CQA) of the product [13].

Currently the primary approach used to determine cell culture media stability is to evaluate many media alternatives in experimental solubility and stability tests. This requires significant manpower and is time consuming and costly. The number of formulations considered, and the parameters tested, could be greatly reduced if we were able to identify and predict the solubility limits for all the species present in media formulations, in silico, to narrow the range of experiments that must be performed. Understanding the thermodynamic properties controlling solubility would represent an important step forward towards designing media that limits precipitation and enables stable media formulations for improved cell culture performance. To achieve this goal, a better understanding and characterization of the intermolecular interactions between the components would be highly beneficial. Since media precipitate often is found to include several amino acid residues, along with other components, we decided to focus our initial efforts on studying interactions between various amino acids before including the effects of salts and other components into the mixture. Fortunately, many of the principles needed for studying nutrient solubilities in silico exist in the field of molecular thermodynamics. These have been used extensively in the chemical and petrochemical industries. Previous studies on the solubility of amino acids in ternary system mixtures consisting of two amino acids (AAs) in water have used Wilson's Model, PC-SAFT, NRTL etc. to model the amino acid activity coefficients and solubilities [14-16]. Currently, determining AA-AA interactions for all possible ternary mixtures is not possible because data on such systems is limited; however, it is not likely that these interaction parameters would be valid or even reasonable in the media or feed formulations because they contain so many different components. Hence, a group contribution thermodynamic model such as UNIFAC can help to address this challenge as it utilizes functional group interaction parameters rather than compound-compound interaction parameters. This is especially advantageous in this case because most amino acids have similar functional groups and many of these functional groups also are present in other media components. In fact, successful efforts in the past have been undertaken to model amino acid solubilities using UNIFAC [17-19]. However, these publications each were restricted to a limited subset of amino acids and present only a small set (less than 5) of ternary system data (two amino acids in water). Furthermore, these studies did not focus on the key amino acids prone to precipitation in cell culture media, perhaps due to difficulties in gathering experimental data on ternary systems of amino acids. Further, none of these studies looked at any of the complexities of cell culture media formulations like pH, the presence of salts, or the presence of other organic compounds like fatty acids or triglycerides.

Therefore, in order to expand the analysis to amino acids prone to precipitation in cell culture media, in-house experimental data and literature experimental data have been used to regress UNIFAC interaction parameters relevant to 15 amino acids that have similar isoelectric points such as leucine and alanine which have pI's within ~1% of each other (supplementary Table S1). For this same reason of maintaining proximal isoelectric points between amino acids, ternary systems of aspartic acid, glutamic acid, histidine, arginine, and lysine were not included in the current study as it is important to model long-range intermolecular electrostatic forces to predict their solubilities and these will be discussed in a forthcoming paper. The isoelectric points of amino acids are shown in Supplementary Table S2.

Unfortunately, the publicly available interaction parameters for original UNIFAC from the Dortmund database does not contain parameters for interactions between the carboxylic acid group and the amine group [20]. This motivated us to undertake our own experimental data collection to address this challenge. Using this experimentally collected data, the resulting regressed set of interaction parameters was used to make predictions on the impact of solubility of an amino acid as a function of concentration of another amino acid. To the author's knowledge, this is the largest set of ternary systems of amino acids modeled in the literature by using one set of interaction parameters. These findings yield significant insights into why precipitation of amino acids such as tyrosine and phenylalanine is observed at concentrations below their own solubility limits. Modeling of the solubilities can not only be used to predict precipitation or reduced solubility limits but also to predict favorable interactions between or among amino acids that can lead to increases in solubility limits. This study also looks at these favorable interactions to provide some insights into increasing the solubility limits of amino acids in water by simply making pots of amino acids that increase each other's solubilities. To demonstrate the applicability beyond ternary systems, predictions of solubility limits in quaternary systems of three amino acids in water also have been explored and recommendations for improving solubilities of amino acid solutions are discussed.

Cell culture media is complex and contains many components. The modeling approach presented here represents an important step in transforming the process of media formulation from an experimentally driven and data intensive process into one that will enable us to computationally predict the interactions in media formulations of the amino acids needed to grow CHO and other mammalian cells.

# 2 Experimental determination of amino acid solubilities

## 2.1 Sample preparation:

All amino acids were purchased from Sigma Aldrich and amino acid solutions were prepared using HPLC-grade water. Experiments were performed to study the solubility of an amino acid (AA1) as a function of the concentration of a second amino acid (AA2). For sample preparation, various amounts of AA2 were weighed out and dissolved in HPLC grade water in 50 mL conical tubes. Multiple samples of AA2 were prepared that span from zero to the solubility of AA2. Then, these samples were placed on a rotator and allowed to equilibrate for at least 24 hours. After dissolving AA2, appropriate amounts of AA1 were measured and added to the tubes containing AA2. An amount of AA1 was added to each sample so that the concentration of AA1 exceeded its solubility limit in pure water. To ensure equilibration, these ternary samples were placed on a rotator for another 48 hours. After equilibration, the AA1 oversaturated solutions were processed through 0.45 μm filters (Millipore) to remove undissolved amino acid particles and to obtain saturated solutions of AA1. The saturated solutions were immediately diluted by 2X to prevent precipitation of the dissolved species due to temperature fluctuations.

## 2.2 HPLC analysis:

Solubility measurements were performed utilizing an automated pre-column derivatization HPLC method. The column used was an Agilent Poroshell HPH C18. Briefly, amino acid samples were first neutralized in borate buffer to neutralize the amino terminus. Subsequently, the primary amines on the amino acids were mixed with ortho-phthaldehyde (OPA). The derivatized product enables the separation of different amino acids with a reverse-phase mode column and it enables sensitive UV detection at 338 nm. Operating conditions were taken from the protocols provided by Agilent.

For each amino acid of interest, a three-point calibration curve was generated within the linear range where the output signals are directly proportional to the concentration of the amino acid in the sample. Before analysis, each amino acid sample was diluted appropriately to obtain a signal measurement within the range of the respective amino acid calibration curve. Each data point then was analyzed with at least three separate sample injections into the HPLC. After calculating the amino acid concentrations of the diluted samples, the solubility limit of the amino acid was obtained from the product of the measured concentration of the diluted sample and the dilution factor used for that sample. A schematic of the experiments is shown in Figure 1.

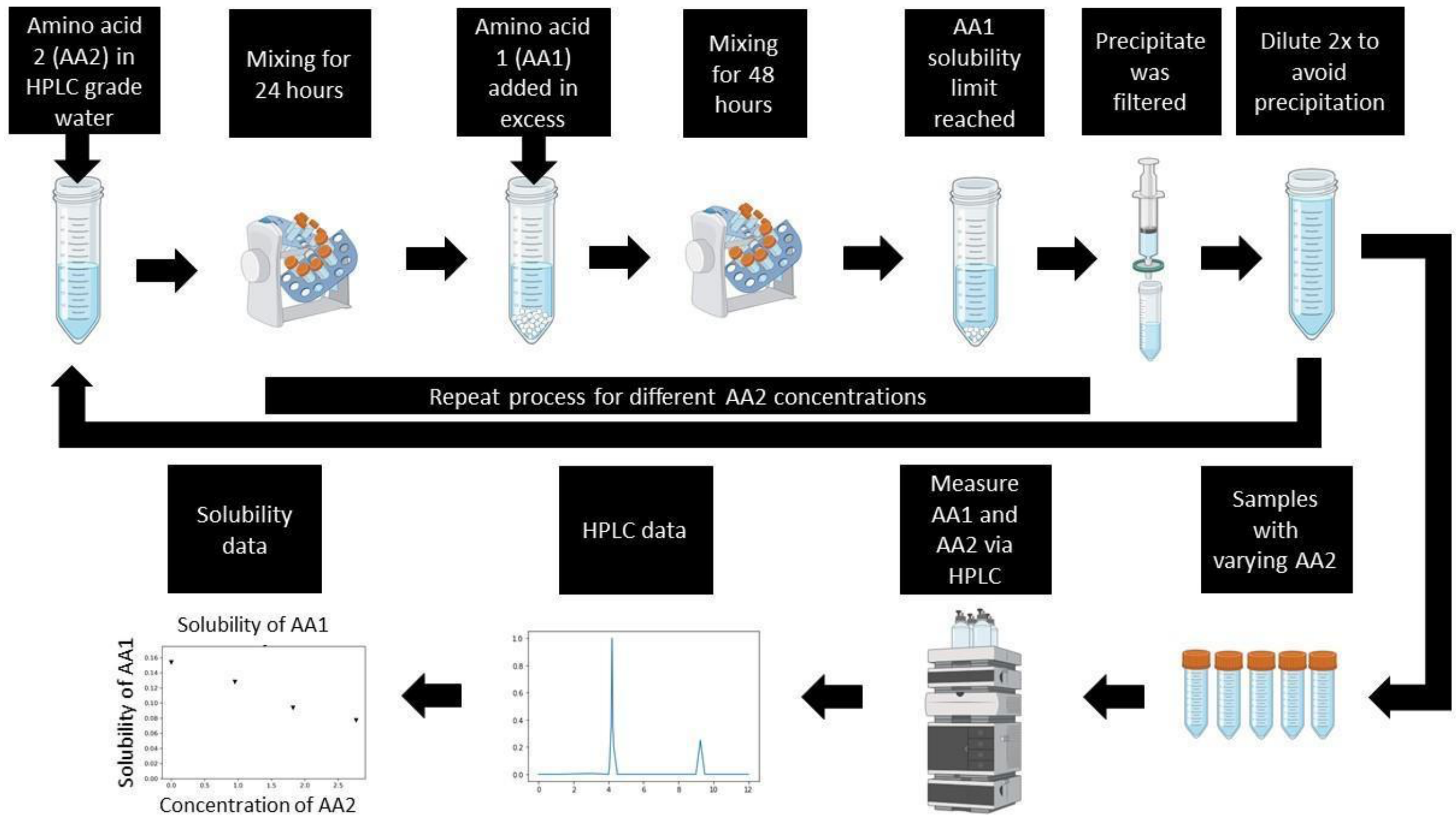

**Figure 1:** Measuring the solubility of an amino acid (AA1) in ternary systems consisting of another amino acid (AA2) in water was performed as described in the schematic. The first step involved dissolving a specific amount of AA2 in water by mixing for 24 hours. The second step involved adding excess AA1 into the solution and mixing for 48 hours. Precipitate was filtered and the solution was diluted 2X to prevent further precipitation while handling the samples. This process was repeated for various concentrations of AA2. The solubility of AA1 was measured by using a HPLC. Thus, the impact of AA2 concentrations on AA1 solubility was experimentally determined.

# 3 Model description

## 3.1 UNIFAC and activity coefficient

Chemical and phase equilibria are determined by the equality of chemical potentials, or equivalently the equality of fugacities. In this case we are talking about solid-liquid phase equilibria. Hence the equations can be written as follows:

$$\mu_i^S = \mu_i^L \tag{1}$$

$$f_i^S = f_i^L \tag{2}$$

Where $\mu_i^S$ is the chemical potential of species $i$ in the solid phase (S) and $\mu_i^L$ is the chemical potential in the liquid phase (L). $f_i^S$ and $f_i^L$ refer to the fugacity of species $i$ in solid (S) and liquid

(L) phases respectively. The equality of fugacities (Equation 2) can be written in terms of the activity coefficients (for liquids), i.e:

$$f_i^{S_0}(z_i\,,T) = x_i * \gamma_i(x_i,T) * f_i^{L_0}(T, x_i = x_i{}^o) \tag{3}$$

Where $f_i^{S_0}(z_i\,,T)$ is the standard state fugacity of species $i$ in solid phase; it is a function of the composition $z$, of species $i$, at temperature T. Here, $x_i$ is the liquid-phase mole fraction of $i$, $\gamma_i(x_i,T)$ is the activity coefficient of $i$, and $f_i^{Lo}(T, x_i = x_i{}^o)$ is the standard state fugacity of species $i$ in the liquid phase. The standard state fugacity usually is defined as that of the pure liquid of species. However, for aqueous solutions of electrolytes, it is common practice to have the standard state fugacity defined as that of the ionic species at infinite dilution in water, i.e. $x_i{}^o = 0$.

The activity coefficient of a solution provides information on deviation from ideal state of the solution and also provides information on attractive and repulsive forces between molecules present in the mixture. The activity coefficient is a function of short-range interactions such as van der Waals forces or long-range interactions such as ionic forces.

UNIFAC is a group-based thermodynamic model that can predict short-range van der Waals interactions between molecules, allowing the calculation of activity. In the UNIFAC model, the activity coefficient is given in terms of a combinatorial contribution considering entropy effects arising from differences in molecular size and shape, and a residual contribution considering energetic interactions among the functional groups in the mixture. So, each chemical/functional group is parametrized by its group volume (R) and group surface area (Q) which are based on physical (structural) measurements, for the combinatorial contribution. The specific group-to-group energetic interactions are parameterized by interaction parameters ($IP_{ij}$) which can be directly estimated from solubility data and activity coefficient data [22]. Larsen's modified version of UNIFAC was used in this work as it has been shown to be reasonable for predicting solubilities of amino acids in water [19,23].

## 3.2 Set of parameters

Calculation of activity coefficients using UNIFAC requires a set of parameters as discussed here. The Bondi volumes R and surface area Q parameters are tabulated in Table 1. Most of these parameters were taken from previous studies [19,20,24,25]. The parameters for the functional group -TriN were calculated by using the method described by Bondi [26]. The functional group -ACHNH, R and Q values could not found in the literature and could not be calculated and have been approximated to be equal to parameters available from modified UNIFAC [25]. All the amino acids of interest and water can be described as a function of the 15 main groups shown in Table 1. The functional group distribution in each amino acid is shown in Table 2. The regression of interaction parameters is described below in Section 3.4.

**Table 1**: Bondi surface area (Q) and volume (R) of different functional groups.

| Number | Main group | Sub Group | R | Q |
|---|---|---|---|---|
| 1 | $H_2O$ | $H_2O$ | 0.92 [20] | 1.4 [20] |
| 2 | Alpha-CH2 | Alpha-CH2 | 0.6744 [19] | 0.54 [19] |
| | | Alpha-CH | 0.4469 [19] | 0.228 [19] |
| 3 | NH2 | NH2 | 0.6948 [19] | 1.15 [19] |
| 4 | COOH | COOH | 1.3013 [20] | 1.224 [20] |
| 5 | Sc-CH3 | Sc-CH3 | 0.9011 [19] | 0.848 [19] |
| | | Sc-CH2 | 0.6744 [19] | 0.54 [19] |
| | | Sc-CH | 0.4469 [19] | 0.228 [19] |
| 6 | OH | OH | 1 [20] | 1.2 [20] |
| 7 | ACOH | ACOH | 0.8952 [20] | 0.68 [20] |
| 8 | ACH | ACH | 0.5313 [20] | 0.4 [20] |
| | | AC | 0.3652 [20] | 0.12 [20] |
| 9 | CONH2 | CONH2 | 1.4515 [24] | 1.248 [24] |
| 10 | CH3S | CH3S | 1.613 [20] | 1.368 [20] |
| 11 | CH2NH | CH2NH | 1.207 [20] | 0.936 [20] |
| 12 | CH2SH | CH2SH | 1.651 [24] | 1.368 [24] |
| 13 | Histidine ring/IMIDAZOL | Histidine ring/IMIDAZOL | 2.026 [20] | 0.868 [20] |
| 14 | TriN | TriN | 2.1354 [a] | 1.8716 [a] |
| 15 | ACHNH | ACHNH | 1.0486 [25] | 0.8076 [25] |

[a] Calculated in this manuscript

**Table 2**: Functional group distribution of amino acids.

| Compound | Functional group distribution |
|---|---|
| $H_2O$ | $H_2O$ |
| Alanine | NH2, COOH, Alpha-CH, Sc-CH3 |
| Asparagine | NH2, COOH, Alpha-CH, Sc-CH2, CONH2 |
| Cysteine | NH2, COOH, Alpha-CH, CH2SH |
| Glutamine | NH2, COOH, Alpha-CH, 2x Sc-CH2, CONH2 |
| Glycine | NH2, COOH, Alpha-CH2 |
| Phenylalanine | NH2, COOH, Alpha-CH, Sc-CH2, AC, 5x ACH |
| isoleucine | NH2, COOH, Alpha-CH, Sc-CH, Sc-CH2, 2x Sc-CH3 |
| Leucine | NH2, COOH, Alpha-CH, Sc-CH, Sc-CH2, 2x Sc-CH3 |
| Methionine | NH2, COOH, Alpha-CH, 2xSc-CH2, CH3S |
| Proline | Alpha-CH, COOH, 3x Sc-CH2, CH2NH |
| Serine | NH2, COOH, Alpha-CH, Sc-CH2, OH |
| Threonine | NH2, COOH, Alpha-CH, Sc-CH, Sc-CH3, OH |
| Tryptophan | NH2, COOH, Alpha-CH, Sc-CH2, 4x ACH, 3x AC, ACHNH |
| Tyrosine | NH2, COOH, Alpha-CH, Sc-CH2, AC, 4x ACH, ACOH |
| Valine | NH2, COOH, Alpha-CH, Sc-CH, 2x Sc-CH3 |
| Histidine | NH2, COOH, Alpha-CH, Sc-CH2, Histidine-ring |
| Arginine | NH2, COOH, Alpha-CH, 3x Sc-CH2, Arginine tri-N group |
| Aspartic acid | NH2, 2x COOH, Alpha-CH, Sc-CH2 |
| Glutamic acid | NH2, 2x COOH, Alpha-CH, 2x Sc-CH2 |
| Lysine | 2x NH2, COOH, Alpha-CH, 4x Sc-CH2, C(NH2)NH |

## 3.3 Prediction of solubility

To predict solubility of a given molecule in solution using the UNIFAC model, the relevant parameters are required. The physical parameters R and Q are available in the literature, but the interaction parameters must be fit from experimental data. Given a molecule (composed of multiple functional groups) in solution, UNIFAC can use interaction parameters to predict the activity and activity coefficient $\gamma_i$ of the molecule $i$ at a given concentration $x_i$ and temperature $T$. The temperature dependence of the ratio standard state fugacities is given by an Antoine-like equation [19].

$$\ln\left(\frac{f_i^{S_0}}{f_i^{L_0}}\right) = \ln(x_i\gamma_i) = A_i - \frac{B_i}{T} + C_i \ln(T) \tag{4}$$

Where, $A_i$, $B_i$, and $C_i$ are constants that are regressed from experimental solubility measurements at various temperatures for each species $i$. However, in this study, for a particular ternary system, the temperature is constant. Hence, we can reduce the above equation to

$$\frac{f_i^{S_0}}{f_i^{L_0}} = x_i\gamma_i(x_i, T) = a_i \quad (5)$$

In Equation 5, $i$ is an amino acid in solution and T is the temperature. The activity ($a_i$) is constant if temperature is constant and the concentration of molecule of $i$ is at the solubility limit. Let us define the activity of species $i$ at the solubility limits of species $i$ as $a_i^{sol}$. So, as another amino acid is added to the solution along with "$i$", if temperature remains the same (as it is in our protocol), then the activity of "$i$" at the solubility limit remains unchanged. The interactions with the additional amino acid added in liquid solution will have no impact on $a_i^{sol}$. The binary system (one amino acid in water) activity coefficient predictions via UNIFAC were used to determine $a_i^{sol}$ values for each amino acid by calculating the activity coefficient at the solubility limit and multiplying this activity coefficient by the mole fraction of amino acid "$i$" in the binary system. Under isothermal conditions, solving for $x_i$ in Equation 5 will yield the solubility point of amino acid "$i$" in a mixture. Equation 5 is nonlinear and is a function of concentration of the amino acid itself. This can be solved iteratively using the Newton-Raphson method to yield the solubility of the amino acid. This is applicable to a system consisting of any number of amino acids, provided that the concentrations of the other amino acids are constant.

## 3.4 Regression of UNIFAC parameters

A database of activity coefficients for binary systems of one amino acid in water was developed from the literature [16,19,27]. Activity coefficients calculated from PC-SAFT predictions of activity coefficients also were added to this database for systems where experimental data were not available [16,28] (depiction in Figure 2). This database contained activity coefficient data for 19 amino acids, and these are presented in Table 3. Many amino acids share similar functional groups. For example, the functional groups in glycine can be divided into NH2, COOH, Alpha-CH2 and the functional groups in serine can be divided into NH2, COOH, Alpha-CH, Sc-CH2, OH. The possible interactions in a binary system of serine in water are NH2 – COOH, NH2 – AlphaCH, NH2 – ScCH2, NH2 – OH, COOH – AlphaCH, COOH – Sc-CH2, COOH – OH, AlphaCH – ScCH2, Alpha-CH – OH, and ScCH2 - OH. The possible interactions in a ternary system of glycine-serine-water are NH2 – COOH, NH2 – AlphaCH, NH2 – ScCH2, NH2 – OH, COOH – AlphaCH, COOH – Sc-CH2, COOH – OH, AlphaCH – ScCH2, Alpha-CH – OH, and ScCH2 - OH. Since the interactions are the same in both cases, if the assumptions of UNIFAC are reasonable and if the data regression process is rigorous, it should be possible to make predictions of ternary system behavior for the glycine-serine-water system from the parameters regressed from the binary system glycine-water and serine-water activity coefficient data only (Figure 2). Many amino acids share similar functional groups, as can be seen in Table 2, so this approach can be used for many cases of binary and ternary interactions. The results and discussion demonstrate that this is indeed a successful approach.

**Table 3**: Binary activity coefficient data.

| Compound in water | Experimental or PC-SAFT |
|---|---|
| Alanine | Literature experimental data [19] |
| Asparagine | PC-SAFT model data [28] |
| Cysteine | Literature experimental data [16] |
| Glutamine | PC-SAFT model data [28] |
| Glycine | Literature experimental data [19] |
| Phenylalanine | PC-SAFT model data [28] |
| Leucine | Literature experimental data [16] |
| Methionine | Literature experimental data [16] |
| Proline | Literature experimental data [19] |
| Serine | Literature experimental data [19] |
| Threonine | Literature experimental data [19] |
| Tryptophan | PC-SAFT model data [28] |
| Tyrosine | PC-SAFT model data [28] |
| Valine | Literature experimental data [19] |
| Histidine | PC-SAFT model data [28] |
| Arginine | Literature experimental data [27] |
| Aspartic acid | PC-SAFT model data [28] |
| Glutamic acid | PC-SAFT model data [28] |
| Lysine | Literature experimental data [27] |

However, this approach falls short when both of the amino acids involved have ‘rare’ groups present in their side chains. An example is methionine and serine. Serine is one of only two amino acids that contain the -OH group, while methionine is the only amino acid that contains the -S-CH3 group. Regressing the interaction parameters from methionine and serine binary (amino acid-water) systems will not provide the interaction parameters between -OH and -S-CH3, as these are never simultaneously present in any binary solution. Thus, the approach of only using binary activity coefficient data will fail to predict ternary system behaviors between methionine and serine. This can be seen in Figure 2. To make ternary system predictions in such systems, we need to determine the interaction parameters between this set of functional groups by regressing the interaction parameters from the ternary system solubility data that is predicted based on the activity coefficient determined via UNIFAC as described in Section 3.3. The approach used for regression of all the interaction parameters used in this study is described below. The Larsen’s UNIFAC model [22] predicts the activity coefficient as a function of the mole fraction of the components in the media. The activity coefficient of chemical $i$ is represented as

$$\gamma_i = f_{Larsen\ UNIFAC}(x_i, x, T, IP) \tag{6}$$

Where $x_i$ is the mole fraction of the chemical $i$, $x$ is a vector of the mole fractions of the other chemicals in the mixture including water, T is the temperature, and *IP* represents the set of interaction parameters among the functional groups present.

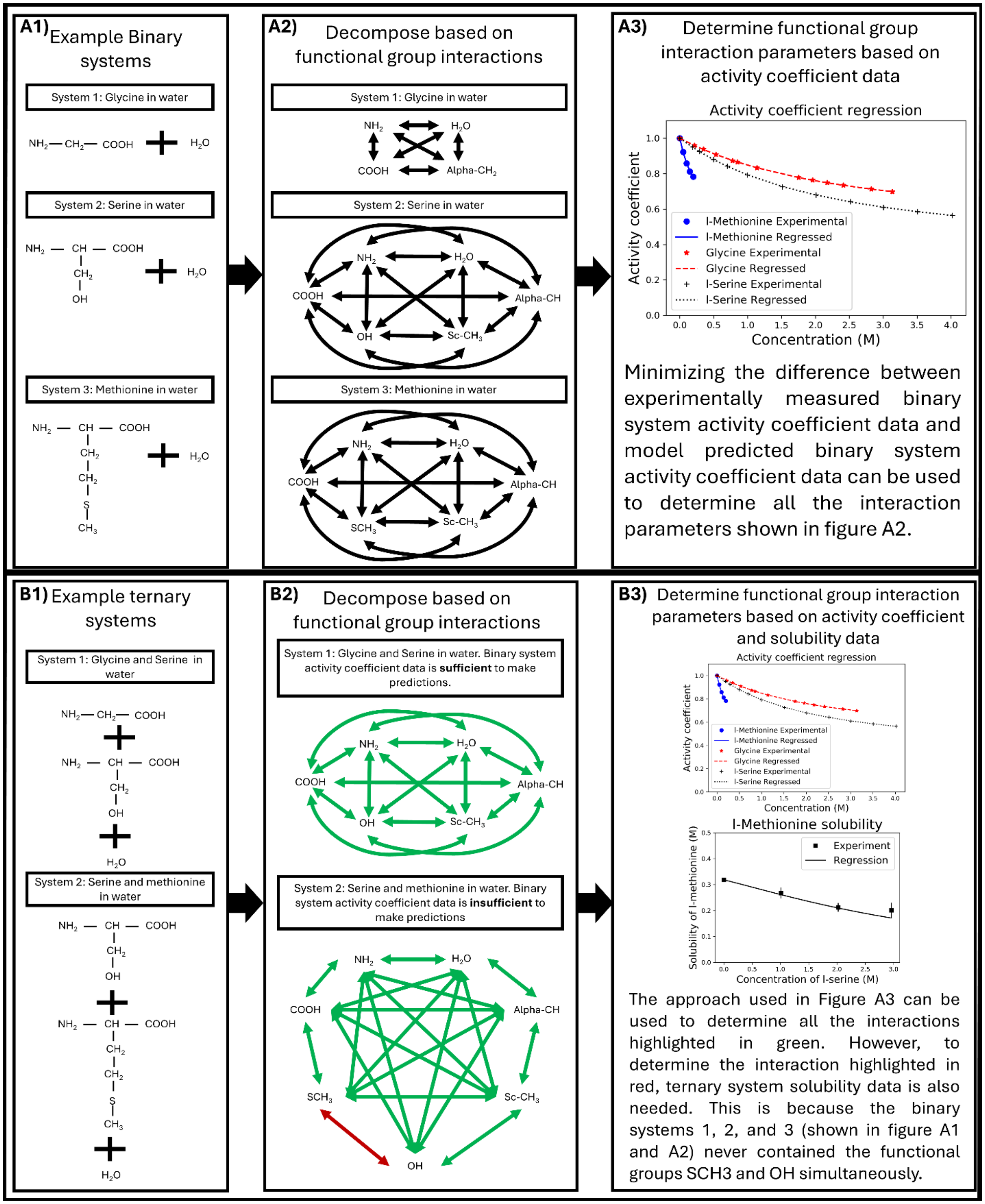

**Figure 2:** The process of regressing UNIFAC interaction parameters from experimentally measured activity coefficient and solubility data is described in this schematic. Examples of binary systems of amino acids in water (Plate A1). Next the interactions between the functional groups in the binary systems shown in Plate A1 are described (Plate A2). The UNIFAC interaction parameters between these groups can be determined by regressing the model onto experimentally measured binary system activity coefficient data as shown in Plate A3. An example ternary system of glycine and serine in water is shown Plate B1. The interaction parameters determined from binary system activity coefficient data in Plate A3 are sufficient to determine all the interactions between functional groups (in ternary system 1) as seen in Plate B2. However, to model a system of serine and methionine in water, the interaction between SCH3 and OH functional groups (highlighted in red) cannot be determined from binary system data alone. Thus, ternary system solubility data needs to be used to determine these interaction parameters (Plate B3).

Experimentally measured solubility data for sixty ternary systems (Table 4) was compiled from literature sources and in-house experiments. Forty (1 to 40 in Table 4) of these ternary systems consisted of amino acids with similar isoelectric points and were used to demonstrate the application of UNIFAC for modeling solubility of amino acids. Determining the activity coefficients of amino acids in the remaining twenty (41 to 60 in Table 4) ternary systems will require modeling long range interactions and will be pursued in a subsequent study.

**Table 4**: Ternary system solubility data at specific temperatures. Data obtained from specific references listed (citations listed as superscript) or from in-house experimental measurements (designated with superscript a).

| Number | Amino acid 1 | Amino acid 2 | Temperature (°C) |
|---|---|---|---|
| 1 | l-serine | proline | 17.5 [a] |
| 2 | | l-methionine | 17 [a] |
| 3 | | l-asparagine | 19.5 [a] |
| 4 | dl-serine | glycine | 25 [29] |
| 5 | | dl-alanine | 25 [19] |
| 6 | l-methionine | l-asparagine | 20 [a] |
| 7 | | l-tryptophan | 20 [a] |
| 8 | | l-serine | 18 [a] |
| 9 | | proline | 21.5 [a] |
| 10 | l-tyrosine | proline | 17 [a] |
| 11 | | l-phenylalanine | 20 [a] |
| 12 | | l-serine | 19 [a] |
| 13 | | glycine | 25 [30] |
| 14 | | l-leucine | 25 [30] |
| 15 | l-tryptophan | l-asparagine | 18 [a] |
| 16 | | l-phenylalanine | 18 [a] |
| 17 | | l-methionine | 21 [a] |
| 18 | | l-serine | 19 [a] |
| 19 | l-valine | l-leucine | 25 [31] |
| 20 | dl-valine | dl-alanine | 25 [19] |
| 21 | dl-alanine | glycine | 25 [29] |
| 22 | | dl-valine | 25 [29] |
| 23 | | dl-serine | 25 [19] |
| 24 | l-alanine | l-leucine | 30 [31] |
| 25 | glycine | dl-serine | 25 [29] |
| 26 | | dl-alanine | 25 [29] |
| 27 | | dl-phenylalanine | 25.05 [32] |
| 28 | | l-leucine | 25 [30] |
| 29 | | l-tyrosine | 25 [30] |
| 30 | dl-phenylalanine | glycine | 25.05 [32] |
| 31 | l-phenylalanine | l-asparagine | 18 [a] |
| 32 | | l-methionine | 18 [a] |
| 33 | | l-tryptophan | 18 [a] |
| 34 | | l-serine | 18 [a] |
| 35 | l-leucine | glycine | 25 [30] |
| 36 | | l-tyrosine | 25 [30] |
| 37 | | l-valine | 25 [31] |
| 38 | | l-alanine | 30 [31] |
| 39 | l-aspartic acid | l-glutamic acid | 25 [33] |
| 40 | l-glutamic acid | l-aspartic acid | 25 [33] |
| 41 | l-serine | l-lysine | 18 [a] |
| 42 | | l-arginine | 18 [a] |
| 43 | l-histidine | l-methionine | 20 [a] |
| 44 | | l-asparagine | 18 [a] |
| 45 | | l-serine | 18 [a] |
| 46 | | proline | 17 [a] |
| 47 | | l-arginine | 22 [a] |
| 48 | l-arginine | l-histidine | 19 [a] |
| 49 | | l-phenylalanine | 17 [a] |
| 50 | | l-asparagine | 17 [a] |
| 51 | | l-lysine | 18.5 [a] |
| 52 | | l-tryptophan | 18.5 [a] |
| 53 | | l-methionine | 18.5 [a] |
| 54 | | l-serine | 18.5 [a] |
| 55 | l-methionine | l-arginine | 18 [a] |
| 56 | l-tyrosine | l-arginine | 17 [a] |
| 57 | l-tryptophan | l-histidine | 17.5 [a] |
| 58 | l-phenylalanine | l-histidine | 17 [a] |
| 59 | l-glutamic acid | glycine | 25 [33] |
| 60 | | l-serine | 25 [33] |

[a] Experimental results published in this work

In this work, the objective function used to determine the optimal set of interaction parameters, defined here as *F(IP)* is a function of the sum of squares of residuals of the activity coefficient data as well as the ternary system solubility data taken collectively.

$$F(IP) = \sum_{i=1}^{n} \sum_{j=1}^{m} \left( \frac{\gamma_{data} - \gamma_{pred}}{\gamma_{data}'} \right)^2 + \sum_{i=1}^{n} \sum_{j=1}^{n-1} \sum_{k=1}^{q} \left( \frac{Solubility_{data} - Solubility_{pred}}{Solubility_{data}} \right)^2 \quad (7)$$

Activity coefficients were calculated using UNIFAC as described above. The ternary system solubilities were calculated by first determining the activity coefficient and subsequently calculating solubility as described in Section 3.3. In Equation 7, $n$ represents the number of amino acids, $m$ represents the number of data points for activity coefficient for each amino acid, $q$ represents the total number of data points for the solubility of each amino acid in the ternary system. $Solubility_{data}$ is the experimentally measured ternary system solubility and $Solubility_{pred}$ is the predicted ternary system solubility. In summary, in this objective function, the first term is the residual of the activity coefficients in binary amino acid systems whereas the second term is the residual of ternary amino acid systems' solubilities. The best set of parameters is determined by minimizing the objective function by using the global optimization algorithm dual annealing with the local optimizer set to Nelder-Mead [34]. The list of interaction parameters determined from binary system activity coefficient data and ternary system solubility data known as the B+T ("Binary and Ternary" dataset) as described in this section is listed in Table 5.

**Table 5**: List of UNIFAC interaction parameters.

| **Main group** | $H_2O$ | Alpha-CH2 | NH2 | COOH | Sc-CH3 | OH | ACOH | ACH | CONH2 | CH3S | CH2NH | CH2SH | Histidine ring/IMIDAZOL | TriN | ACHNH |
|---|---|---|---|---|---|---|---|---|---|---|---|---|---|---|---|
| $H_2O$ | 0 | -70.95 | 813.31 | -392.64 | 615.06 | 502.45 | 2102.38 | 933.95 | 502.29 | 1469.47 | 122.32 | 280.85 | 147.83 | 173.16 | 1475.08 |
| Alpha-CH2 | -497.25 | 0 | 1520.36 | -1166.06 | 794.66 | 96.22 | 1125.42 | -44.42 | -7.11 | 1464.63 | 62.64 | -5.18 | -17.51 | -7.96 | 852.02 |
| NH2 | -237.36 | 695.04 | 0 | 914.84 | 1292.93 | 314.70 | 1850.05 | 856.91 | -23.16 | 1887.09 | 320.38 | -85.99 | -45.07 | -33.05 | 1225.26 |
| COOH | 165.96 | -463.055 | 1765.77 | 0 | 1462.47 | 418.97 | 1867.49 | 1126.01 | -22.46 | 1903.64 | 195.23 | -104.28 | -52.10 | -32.33 | 1241.16 |
| Sc-CH3 | -345.22 | 781.38 | 364.96 | 971.03 | 0 | 434.40 | 737.31 | 1005.70 | -414.05 | 1760.99 | -610.26 | NA | -24.62 | -46.00 | 1068.28 |
| OH | -199.35 | -744.21 | 1218.64 | -1049.93 | 1255.23 | 0 | 1843.03 | 649.55 | -21.92 | 492.41 | -781.40 | NA | NA | NA | 1194.14 |
| ACOH | -923.34 | -1656.55 | 526.42 | 90.16 | 225.41 | 682.37 | 0 | 804.70 | NA | NA | 253.33 | NA | NA | NA | NA |
| ACH | -579.31 | -1011.26 | 854.92 | -1252.21 | -319.00 | 641.15 | 2044.54 | 0 | 34.71 | 1936.60 | 397.64 | NA | NA | NA | 635.68 |
| CONH2 | -533.76 | -15.10 | 17.21 | -24.39 | -444.71 | -797.09 | NA | -117.84 | 0 | 514.60 | NA | NA | NA | NA | -249.29 |
| CH3S | -581.21 | 366.13 | 1270.90 | 463.76 | 1013.65 | -526.09 | NA | 1235.97 | 172.83 | 0 | -1043.73 | NA | NA | NA | 726.00 |
| CH2NH | -454.44 | -825.03 | 781.24 | -1123.69 | -18.49 | 316.33 | 40.09 | 965.04 | NA | 73.16 | 0 | NA | NA | NA | NA |
| CH2SH | 210.21 | -12.41 | 274.35 | -37.15 | NA | NA | NA | NA | NA | NA | NA | 0 | NA | NA | NA |
| Histidine ring/IMIDAZOL | 102.76 | -0.0056 | -52.06 | -9.91 | -86.99 | NA | NA | NA | NA | NA | NA | NA | 0 | NA | NA |
| TriN | 406.61 | -111.29 | 842.66 | -759.83 | 528.54 | NA | NA | NA | NA | NA | NA | NA | NA | 0 | NA |
| ACHNH | -1006.12 | 97.72 | 854.72 | 140.66 | -1148.72 | 561.81 | NA | -803.34 | 87.95 | -354.46 | NA | NA | NA | NA | 0 |

# 4 Results and discussion

## 4.1 UNIFAC predictions using only binary system data

To demonstrate the applicability of UNIFAC to predict ternary system solubilities, we regressed a set of UNIFAC interaction parameters, as described in Section 3.4, from binary system activity coefficient data of 19 amino acids. Given that these were considering only binary data, we were able to calculate the activity coefficients directly from data using Equation 7, without using the second ‘solubility’ term of the equation. This set of interaction parameters known as the B (‘Binary’ dataset) parameters is available in the supplementary Table S2. The regression of activity coefficients from the binary system data alone is shown in Figure 3. Figure 3 includes amino acids with varying solubility limits and different interactions with water, serine and tryptophan being examples of very soluble and sparsely soluble amino acids respectively. This figure shows that phenylalanine, tryptophan, leucine, valine, and proline have activity coefficients of greater than 1. This indicates “repulsive” forces between water and the amino acid. More precisely, the term “repulsive” used here refers to systems where the *ij* interactions are less than the arithmetic mean of the *ii* and *jj* interactions. This is common for van der Waals interactions because for weak short-range interactions, the *ij* interaction typically is close to the geometric mean of the *ii* and *jj* interactions and the geometric mean always is less than the arithmetic mean.

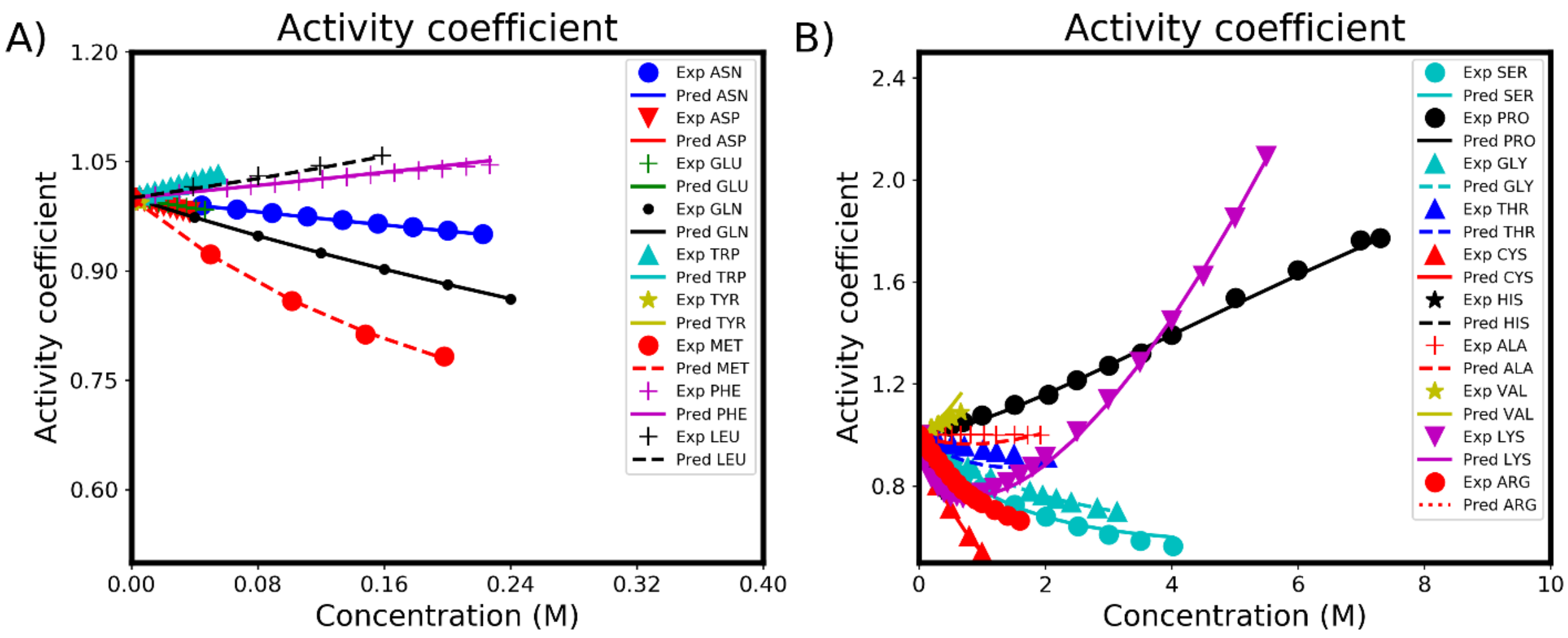

**Figure 3:** Activity coefficients of amino acids with low solubilities are plotted in Plate 3A and the activity coefficients of amino acids with high solubilities are plotted in Plate 3B. UNIFAC interaction parameters (B interaction parameters; Supplementary Table S2) were regressed from binary system activity coefficient data (Table 3) from the literature.

Amino acids methionine, tyrosine, glutamine, asparagine, glutamic acid, aspartic acid, alanine, threonine, serine, glycine, arginine, cysteine and histidine have activity coefficients below 1 at low concentrations. This indicates attractive forces between water and the amino acid. Through optimization of the objective function, the set of UNIFAC binary interaction parameters ("B" parameters) were successfully regressed to describe a diverse set of interactions between amino acids and water with high accuracy.

Moreover, we were encouraged to see that this dataset of B interaction parameters (Supplementary Table S2) that was determined solely from binary system data was able to make solubility predictions of amino acids in ternary systems. The predictions of ternary system solubility from binary system activity coefficient data are shown in Figure 4. The predictive power of this model is highlighted by the fact that these predictions were made with a total error of 8%. For example, the addition of dl-alanine to a solution of dl-valine and addition of l-alanine to a solution of l-leucine resulted in decreasing solubilities. Examples of increased solubility in the dataset included the addition of glycine to a solution of dl-serine, glycine to a solution of dl-phenylalanine, dl-serine to a solution of glycine. As seen by the agreement between the model predictions and the experimental measurements in Figure 4, both types of effects in ternary systems were successfully predicted from a model that was trained only on binary system activity coefficient data. These examples highlight that this functional group approach is effective in predicting different interactions between amino acids simply from binary amino acid datasets. These findings also demonstrate that ternary system predictions can be made from binary system data alone and indicate that it may extend to multi-component systems such as cell culture media formulations. However, it must be noted that the predictions in Figure 4 were performed for systems of amino acids with a similar functional group compositions. This is the benefit of using a functional group-based approach such as UNIFAC, where many groups are shared and therefore even small datasets of amino acids in water contain enough information to create accurate predictions. However, as mentioned previously, using binary system data to make predictions for ternary systems that do not have similar functional groups can be challenging. The regression process is described in Section 3.4 and highlighted in Figure 2. Two model systems were used to demonstrate this point: serine-glycine (similar group composition) and serine-methionine (not many coinciding groups). The UNIFAC interaction parameters between the different functional groups (H2O, AlphaCH2, NH2, COOH, ScCH3, SCH3, OH) used to represent serine, glycine, and methionine in water highlight these challenges. In this case, the UNIFAC model with the B interaction parameters (Supplementary Table S2) regressed solely from binary system data was used to calculate the solubility of dl-serine as a function of concentration of glycine, and model predictions were found to be in very good agreement with the experimental data. This prediction of ternary system behavior from binary system activity coefficient data was feasible because all the interaction parameters between the functional groups (H2O, AlphaCH2, NH2, COOH, ScCH3,

OH) used to model dl-serine and glycine systems could be estimated from binary system data alone. However, the prediction for ternary system of l-serine in l-methionine is not feasible as seen in Figure 2, the interaction parameter between groups -SCH3 and -OH cannot be estimated from binary systems data alone. This is because, unlike the serine-glycine example where the binary dataset contained examples of these functional groups interacting in water, the -SCH3 to -OH functional group does not exist in a dataset of only one amino acid in water. Ternary solubility data will be required to estimate these interaction parameters. However, the minimum number of ternary systems required to characterize interactions between the 'rare' side chain groups is still lower (8 rare side chain groups, 28 ternary systems required (${}^{8}C_{2}$)) compared to an activity model that relies on component or compound interactions rather than group interactions. In the case of 15 amino acids as in this paper, it would take 105 (${}^{15}C_{2}$) experimental datasets to fully characterize all ternary interactions. In addition, unlike our method, this approach would be guaranteed not scale to larger systems (e.g., quaternary, discussed more in Section 4.4). Thus, UNIFAC still requires a much smaller minimal dataset despite this limitation. In order to improve the predictions, it is necessary to regress a set of interaction parameters from both binary systems and ternary systems simultaneously. The B+T interaction parameters (Table 5) were used to model ternary system solubilities as shown in Figure 6 as described in Section 3.4.

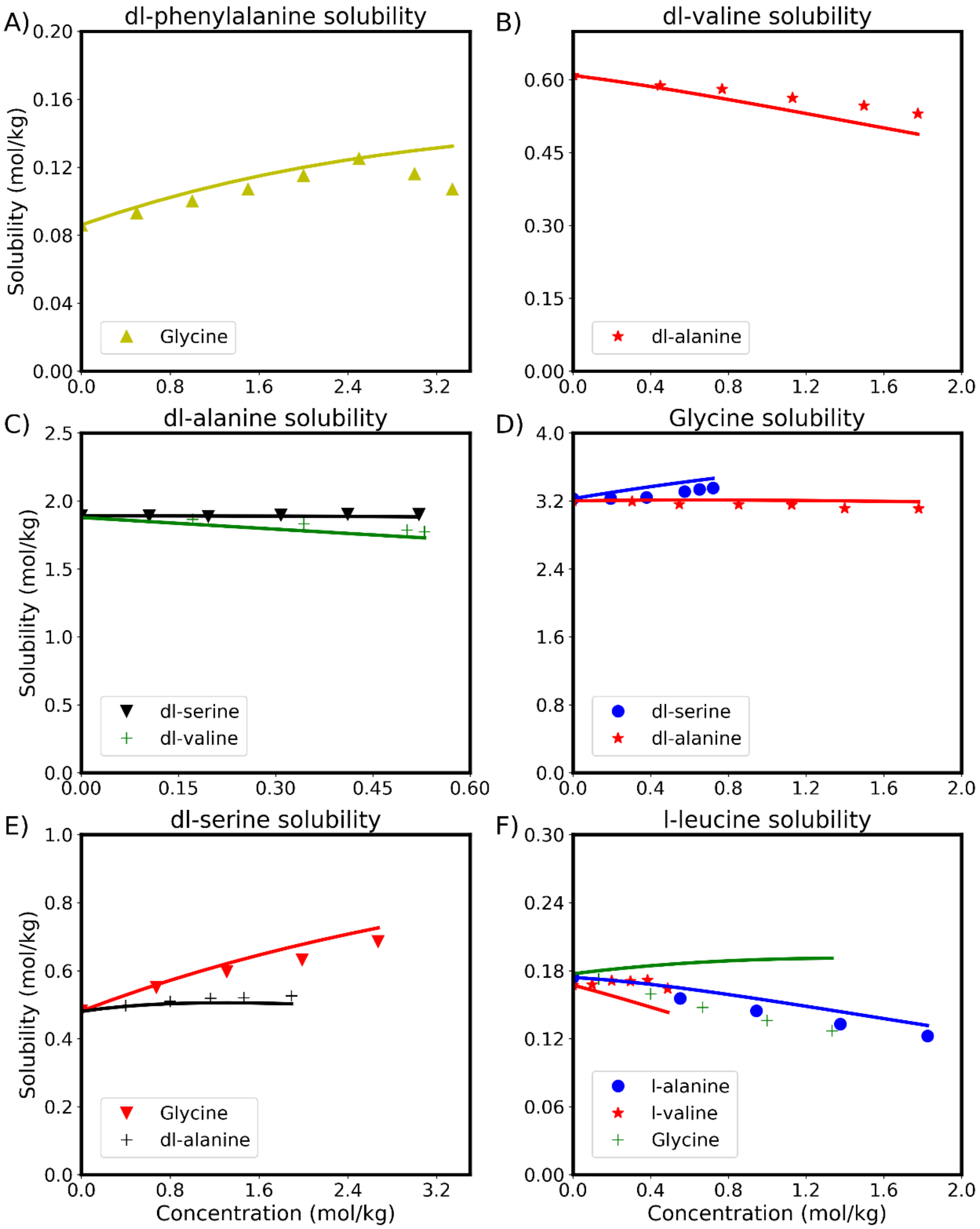

**Figure 4:** Solubility predictions of ternary systems of two amino acids in water by using UNIFAC interaction parameters (B interaction parameters; Supplementary Table S2) determined from only binary system activity coefficient data. A) Solubility of dl-phenylalanine in system of glycine in water. B) Solubility of dl-valine in system of dl-alanine in water. C) Solubility of dl-alanine in systems of dl-serine in water and dl-valine in water. D) Solubility of glycine in system of dl-serine in water and dl-alanine in water. E) Solubility of dl-serine in systems of glycine in water and dl-

alanine in water. F) Solubility of l-leucine in systems of l-alanine in water, l-valine in water, and glycine in water.

## 4.2 Regression and prediction of ternary system

A database consisting of 40 ternary systems was curated based on experimental data generated as part of this study as well as data from previous studies [19,29-33]. Although it would be ideal to have a dataset with activity coefficient data for all components in ternary solution, approaches for measuring activity coefficients of solutes in aqueous solutions rely on measuring the properties of the solvent and then using the Gibbs-Duhem equation to estimate the activity of the solute. Since the solvent is affected by all of the solutes in a multi-component solution, it is not possible to differentiate between the contributions of each solute and thus the activity of individual solutes in a mixed solution cannot be estimated. So, as mentioned in Section 3.4, experimental solubility data, is used to estimate parameters via the UNIFAC activity model. The data from the literature mostly consisted of fairly simple amino acids such as glycine, serine, alanine, valine, and leucine. Glycine, serine, alanine, and valine do not typically have solubility issues in cell culture media. However, these amino acids could potentially influence the solubility of other amino acids. It was difficult to find ternary system literature data on amino acids with complex side chains such as tyrosine, phenylalanine, methionine, etc. Yet, these amino acids are more important for our study as they were detected in cell culture media precipitates due to their low solubility [12]. Hence, in-house experimental data collection was focused towards studying the solubilities of amino acids with complex side chains as a function of concentration of other amino acids. To our knowledge, this is the largest set of ternary system amino acid solubility data reported. Studies in the literature do not go beyond 8 ternary systems [30,31]. The curated dataset of ternary systems of amino acids from both in-house measurements and literature, along with the temperatures at which the data was collected are shown in Table 4. A table with all the raw data is available in the Supplementary Material.

Regression of the interaction parameters from all the activity coefficient data and ternary system solubility data as described in Section 3.4 yielded the B+T set of interaction parameters shown in Table 5. This set of interaction parameters represent 19 binary amino acids datasets (activity coefficients) and 40 ternary amino acid datasets (solubilities) simultaneously regressed and includes data for all the interactions among functional groups in this dataset. This set of interaction parameters (Table 5) was used to model ternary system solubilities; the results are shown below. As a first check of the validity of the parameter set, we tested this newly fit data set on its ability to predict binary amino acid system activity coefficients. For the vast majority of amino acids, there were no significant differences in the model predictions of binary system activity coefficients as seen in Figure 5. So, despite the addition of a new, more complex, and independent dataset, the predictions of binary system activity coefficients are still in agreement with experimental data and capable of predicting a diverse set of interactions among amino acids and water with remarkable accuracy. The only amino acids that showed changes in the model fit of activity coefficient after being regressed with the ternary dataset are tyrosine and methionine. This could partly be explained by the large differences in solubility of tyrosine or methionine (or its impact on solubility of other amino acids) as a function of the other amino acids and that the activity coefficient "data" was taken from literature PC-SAFT model predictions [28]. While the binary data for tyrosine came from activity coefficients from Do's model [28], the experimental data we collected for ternary systems came from in-house solubility measurements via HPLC, as denoted in Section 2.2. So, this data is subject to sensitivity challenges due to the exceptionally low solubility of tyrosine. This is an inherent challenge of leveraging data from different methods and literature sources and will be explored later in this paper. Nevertheless, the tyrosine and methionine binary system activity coefficient fit is still in qualitative agreement with the PC-SAFT calculations or experimental measurements.

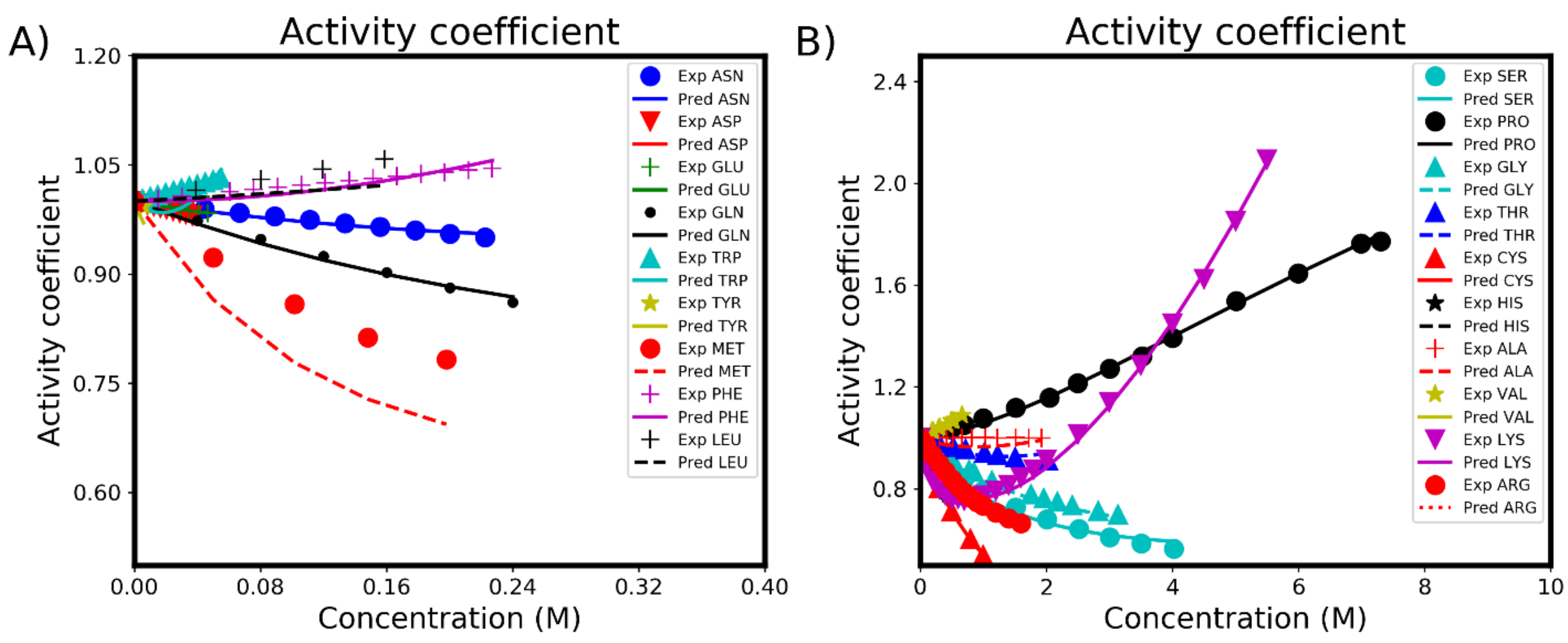

**Figure 5:** Activity coefficients of amino acids with low solubilities are plotted in Plate 5A and the activity coefficients of amino acids with high solubilities are plotted in Plate 5B. Activity coefficient values were plotted after regression of UNIFAC interaction parameters (B+T interaction parameters; Table 5) to both binary system activity coefficient data (Table 3) and ternary system solubility data (Table 4) via Equation 7.

After confirmation that the activity coefficient predictions were not drastically altered by inclusion of ternary system solubility data in the regression procedure, we proceeded to use the B+T UNIFAC interaction parameters shown in Table 5 to model solubilities involving the ternary system data shown in Table 4. Prior to discussing the quality of the predictions, it is necessary to develop some metrics to highlight and quantify the quality of fit for such a large data set. These metrics are described below.

The normalized individual error has been defined as follows:

$$Normalized\ individual\ error = \frac{\sum_{i=1}^{number\ of\ experiments} \left| \frac{Predicted\ ternary\ solubility - Measured\ ternary\ solubility}{Measured\ binary\ solubility} \right| \times 100}{number\ of\ experiments} \tag{8}$$

The ternary difference has been defined as follows:

$$Ternary\ difference = \frac{\sum_{i=1}^{number\ of\ experiments} \left| \frac{Measured\ binary\ solubility - Measured\ ternary\ system\ solubility}{Measured\ binary\ Solubility} \right| \times 100}{number\ of\ experiments} \tag{9}$$

The experimental metric and effect have been defined as follows:

$$Experimental\ metric = \frac{Finalexperimentalternarysolubility - Binary\ solubility}{Binary\ solubility} \tag{10}$$

$$\begin{aligned} Experimental\ effect = \{ & Unchanged\ if\ |Experimental\ metric| < 0.05 \\ & Increased\ if\ Experimental\ metric < -0.05 \\ & Decreased\ if\ Experimental\ metric > 0.05\} \end{aligned} \tag{11}$$

The predicted effect has been defined as follows:

$$Predicted\ metric = \frac{Finalpredictedternarysolubility - Binary\ solubility}{Binary\ solubility} \tag{12}$$

$$\begin{aligned} Predicted\ effect = \{ & Unchanged\ if\ |Predicted\ metric| < 0.05 \\ & Increased\ if\ Predicted\ metric < -0.05 \\ & Decreased\ if\ Predicted\ metric > 0.05\} \end{aligned} \tag{13}$$

The metrics defined in Equations 8, 9, and 10 were normalized by binary system solubility and number of experiments due to the nature of the dataset. Amino acids have vastly different solubilities, and some amino acids had more datapoints in this dataset. So, to allow comparison between the prediction performance of different amino acids and to have an overall performance metric of our model, it was important to normalize the errors. The *normalized individual error* metric denoted the normalized difference in model prediction and experimental datapoint, a measure of the model's accuracy. The '*ternary difference*' metric is a measure of the deviation of solubility of an amino acid alone in water, as compared to it being in the presence of another amino acid and water. By quantifying the divergence from binary amino acid behavior, this metric shows which amino acid pairings benefit the most from using our ternary dataset approach to accurately predict solubility. Additionally, we observed 3 classes of ternary system behaviors that we sought to define qualitatively. These behaviors were whether the addition of an amino acid increased, decreased, or did not affect the ternary system solubility of an amino acid while compared to its binary system solubility. Equations 12 and 13 express the behavior qualitatively by using a threshold of 5% (same order of magnitude as operator or instrument measurement error).

The ternary system regression results obtained from regression of the interaction parameters to both binary system activity coefficient data and the ternary system solubility data (Equation 7) are shown in Figure 6. This may be the first study to regress a model to the solubilities of amino acids in ternary systems at this scale. The average value of the normalized individual error (defined in Equation 8 as the difference between model predicted ternary system solubility and the experimentally measured ternary system solubility divided by the experimentally measured binary system solubility) between predictions and experimental data for 40 ternary systems (Figure 6 and Table 6) is approximately 3% and the average value of the ternary difference (defined in Equation 9 as the difference between the experimentally measured binary system solubility and the experimentally measured ternary system solubility divided by the experimentally measured binary system solubility) for the 40 ternary systems is 12%. Hence, if the binary solubility is assumed to be constant the average deviation between the experimentally measured ternary system solubility and the experimentally measured binary system solubility is 12% (Equation 9) and the average deviation between the experimentally measured ternary system solubility and the model predicted ternary system solubility is 3% (Equation 8). The context behind these numbers can be highlighted with an example. Tyrosine is an amino acid that is crucial for consistent cell culture performance. Poor supply of tyrosine can cause reductions in cell growth, mAb titer production, and product quality issues such as amino acid misincorporation. For the ternary system of tyrosine and serine in water, if we make the assumption that the solubility of tyrosine in the ternary system will not deviate from its binary system solubility, then the ternary difference (Equation 9) value of 75.4 % tells us that this assumption is incorrect. The high ternary difference (Equation 9) value indicates that the solubility of tyrosine in this ternary system deviates drastically from its binary system solubility as observed in Figure 6E. Thus, if this solution is used in cell culture applications, there is a high likelihood of insufficient tyrosine delivery to the cells due to low solubility, which is not reflected in binary system solubility measurements. This is precisely the type of situation for which a mathematical model can be used to predict and screen to avoid precipitation of crucial nutrients. The individual error (defined in Equation 8) of model prediction of tyrosine solubilities as a function of serine concentration is only 8.8 %, indicating that the model is capable of screening tyrosine depletion. Comparison of the individual error (Equation 8) of the model predictions with the ternary difference metric (Equation 9) shows the value of the model and its ability to predict solubilities in ternary systems that result in a large decrease or increase in solubilities. Of the 40 ternary systems, our fit model obtained a prediction error (Equation 8) of less than 5% in 36 of them (the aspartic acid-glutamic acid system had a normalized individual error of 5.5%). This close alignment between predictions and experiment for such a large number of ternary systems underscores the robustness of our data set. Furthermore, the high degree of fitness suggests that the amino acids were sufficiently represented by the functional groups we chose to use and apply in this model. Additionally, the interactive forces being modeled by UNIFAC, which represent short-range van der Waals forces, are ample enough to describe the majority of interactions between these amino acids across a wide range of concentrations: from low to saturation point.

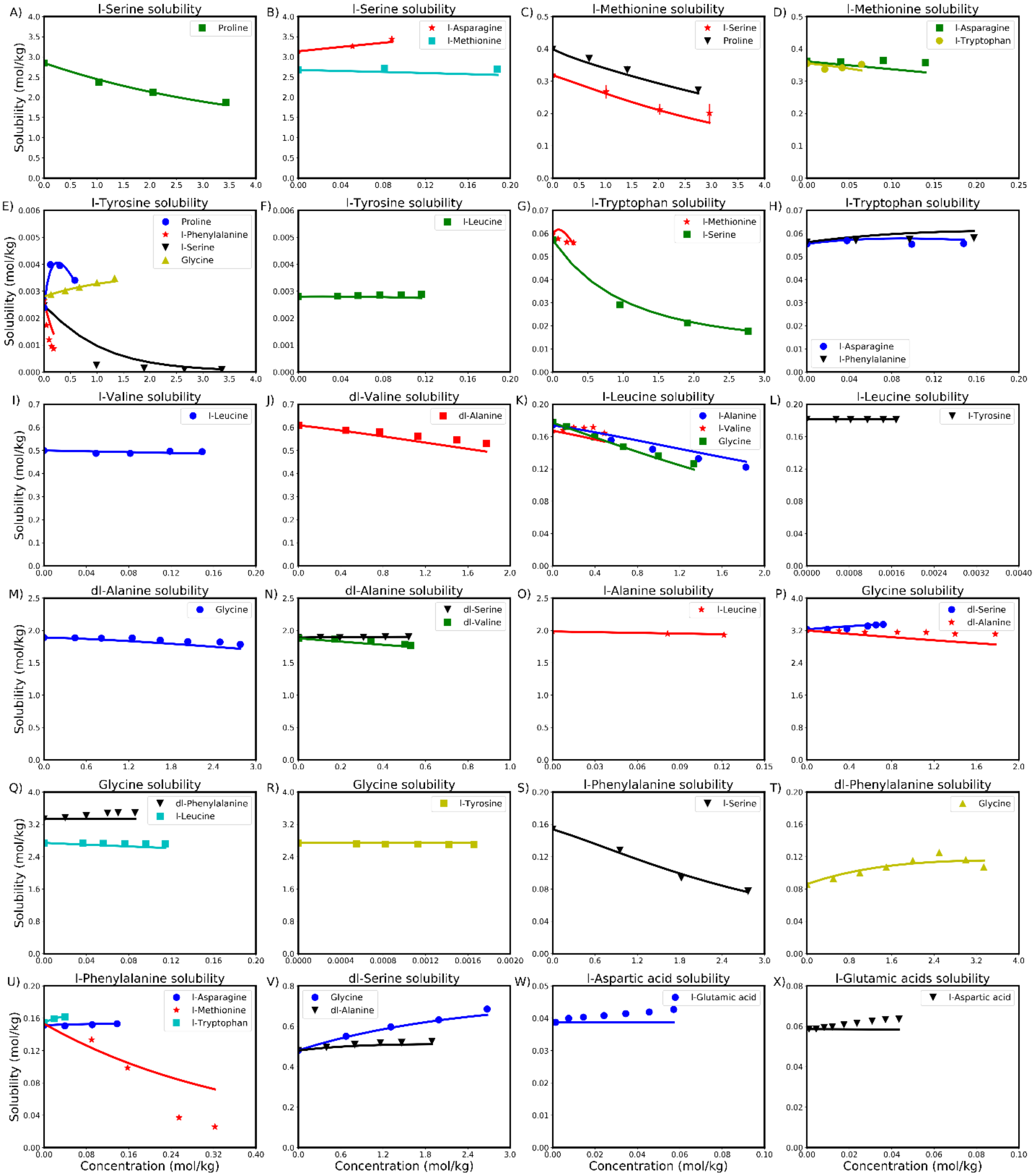

**Figure 6:** Prediction of solubilities of amino acids in 40 different ternary systems by using UNIFAC interaction parameters determined from binary system activity coefficient data and ternary system solubility data. A) Solubility of l-serine in system of proline in water. B) Solubility of l-serine in system of l-asparagine in water and l-methionine in water. C) Solubility of l-methionine in system of l-serine in water and proline in water. D) Solubility of l-methionine in system of l-asparagine in water and l-tryptophan in water. E) Solubility of tyrosine in system of proline in water, l-phenylalanine in water, l-serine in water, and glycine in water. F) Solubility of l-tyrosine in system of l-leucine in water. G) Solubility of l-tryptophan in system of l-methionine in water and l-serine in water. H) Solubility of l-tryptophan in system of l-asparagine in water and l-phenylalanine in water. I) Solubility of l-valine in system of l-leucine in water. J) Solubility of dl-valine in system of dl-alanine in water. K) Solubility of l-leucine in system of l-alanine in water, l-valine in water, and glycine in water. L) Solubility of l-leucine in system of l-tyrosine in water. M) Solubility of dl-alanine in system of glycine in water. N) Solubility of dl-alanine in system of dl-serine in water and dl-valine in water. O) Solubility of l-alanine in system of l-leucine in water. P) Solubility of glycine in system of dl-serine in water and dl-alanine in water. Q) Solubility of glycine in system of dl-phenylalanine in water and l-leucine in water. R) Solubility of glycine in system of l-tyrosine in water. S) Solubility of l-phenylalanine in system of l-serine in water. T) Solubility of dl-phenylalanine in system of glycine in water. U) Solubility of l-phenylalanine in system of l-asparagine in water, l-methionine in water, and l-tryptophan in water. V) Solubility of dl-serine in system of glycine in water and dl-alanine in water. W) Solubility of l-aspartic acid in system of glutamic acid in water. X) Solubility of l-glutamic acid in system of l-aspartic acid in water.

**Table 6**: Comparison of predictions to experimental data.

| Number | Amino acid 1 | Amino acid 2 | Normalized Individual error % | Ternary difference % | Experimental effect | Predicted effect |
|---|---|---|---|---|---|---|
| 1 | l-serine | Proline | 1.2 | 19.1 | Decreased | Decreased |
| 2 | | l-methionine | 3.0 | 0.8 | Unchanged | Unchanged |
| 3 | | l-asparagine | 0.9 | 4.5 | Increased | Increased |
| 4 | dl-serine | Glycine | 1.7 | 22.6 | Increased | Increased |
| 5 | | dl-alanine | 1.5 | 5.9 | Increased | Increased |
| 6 | l-methionine | l-asparagine | 4.3 | 0.6 | Unchanged | Decreased |
| 7 | | l-tryptophan | 2.2 | 2.4 | Unchanged | Unchanged |
| 8 | | l-serine | 3.0 | 21.6 | Decreased | Decreased |
| 9 | | Proline | 2.6 | 13.7 | Decreased | Decreased |
| 10 | l-tyrosine | Proline | 3.9 | 43.7 | Increased | Increased |
| 11 | | l-phenylalanine | 18.7 | 44.2 | Decreased | Decreased |
| 12 | | l-serine | 8.9 | 75.4 | Decreased | Decreased |
| 13 | | Glycine | 1.2 | 10.8 | Increased | Increased |
| 14 | | l-leucine | 2.2 | 1.6 | Unchanged | Unchanged |
| 15 | l-tryptophan | l-asparagine | 1.9 | 0.8 | Unchanged | Unchanged |
| 16 | | l-phenylalanine | 3.3 | 1.6 | Unchanged | Increased |
| 17 | | l-methionine | 3.2 | 3.6 | Decreased | Decreased |
| 18 | | l-serine | 1.5 | 45.2 | Decreased | Decreased |
| 19 | l-valine | l-leucine | 1.2 | 1.3 | Unchanged | Unchanged |
| 20 | dl-valine | dl-alanine | 2.7 | 5.6 | Decreased | Decreased |
| 21 | dl-alanine | Glycine | 2.2 | 1.9 | Decreased | Decreased |
| 22 | | dl-valine | 1.1 | 2.3 | Decreased | Decreased |
| 23 | | dl-serine | 0.2 | 0.3 | Unchanged | Unchanged |
| 24 | l-alanine | l-leucine | 0.3 | 1.5 | Unchanged | Unchanged |
| 25 | Glycine | dl-serine | 0.6 | 1.8 | Unchanged | Unchanged |
| 26 | | dl-alanine | 4.0 | 1.4 | Unchanged | Decreased |
| 27 | | dl-phenylalanine | 2.7 | 2.8 | Unchanged | Unchanged |
| 28 | | l-leucine | 2.0 | 0.5 | Unchanged | Unchanged |
| 29 | | l-tyrosine | 1.0 | 1.0 | Unchanged | Unchanged |
| 30 | dl-phenylalanine | Glycine | 4.5 | 23.4 | Increased | Increased |
| 31 | l-phenylalanine | l-asparagine | 0.4 | 0.5 | Unchanged | Unchanged |
| 32 | | l-methionine | 14.3 | 41.6 | Decreased | Decreased |
| 33 | | l-tryptophan | 0.2 | 2.6 | Increased | Increased |
| 34 | | l-serine | 1.6 | 26.4 | Decreased | Decreased |
| 35 | l-leucine | Glycine | 1.2 | 13.6 | Decreased | Decreased |
| 36 | | l-tyrosine | 0.1 | 0.2 | Unchanged | Unchanged |
| 37 | | l-valine | 4.8 | 1.5 | Unchanged | Decreased |
| 38 | | l-alanine | 3.1 | 16.2 | Decreased | Decreased |
| 39 | l-aspartic acid | l-glutamic acid | 5.5 | 5.4 | Increased | Unchanged |
| 40 | l-glutamic acid | l-aspartic acid | 4.1 | 4.0 | Increased | Unchanged |

In some of the ternary systems, the solubility of an amino acid was unaffected by the solubility of another amino acid. On the other hand, the solubility of amino acids in other systems were significantly affected by the presence of another amino acid. This is most readily observable through the '*Ternary difference metric*' (Equation 9) which denotes the difference between the solubility of an amino acid when it is alone in water vs when it is in the presence of another amino acid. As an example, tyrosine's solubility is changed (in this case reduced) by 75.4 % (via *ternary difference metric described in Equation 9*) experimentally, in the presence of serine compared to when it is alone in water, whereas it is virtually unchanged in the presence of leucine. This is due to the intermolecular forces between compounds which push each other into or out of solution. Through UNIFAC, we can accurately model the activity coefficients of amino acids in solution, which can be seen as a low *individual error value (Equation 8)*. The *ternary difference metric (Equation 9)* serves as a quantitative measure of the effects of amino acid interactions on their solubilities, and large values show cases where assuming binary solubilities does not hold but are often ignored in cell media preparation scenarios. Moreover, cases where the *ternary difference metric* (Equation 9) is large, and the *normalized individual metric* (Equation 8) is small denote scenarios where our model accurately predicts solubilities in formulations that have high solubility deviations due to activity effects. These cases show the unique ability of our model to predict viable media formulations, which is not possible in other current media formulation strategies.

More broadly, our model accurately predicts the behavior of interacting amino acids in solution whether they lead to an increase in solubility, decrease in solubility or no change in solubility qualitatively and quantitatively. This flexibility and deftness illustrates that this method represents is a good approach to describe the interactions among amino acids. As a result, our model provides a worthwhile framework for understanding which amino acids help increase the solubility limits of other amino acids and which ones decrease the solubility limit of other amino acids, yielding valuable insights into media development for biomanufacturing. Consequently, the B+T interaction parameters from Table 5 enable us to make predictions for the ternary systems shown in Table 7. The ones in green are the ones that we have experimental data for.

**Table 7:** Potential possible predictions of ternary system. Each element of this matrix represents the solubility prediction for the amino acid in the row as a function of the amino acid in the column. P is used for predictions that can be made with the interaction parameters (Table 5) that we have and N represents predictions that cannot be made based on the interaction parameters (Table 5) that we have. The green boxes represent systems that we have experimental data for.

| AA | GLY | ALA | VAL | CYS | PRO | LEU | MET | TRP | PHE | SER | THR | TYR | ASN | GLN |
|---|---|---|---|---|---|---|---|---|---|---|---|---|---|---|
| GLY | 0 | P | P | P | P | P | P | P | P | P | P | P | P | P |
| ALA | P | 0 | P | N | P | P | P | P | P | P | P | P | P | P |
| VAL | P | P | 0 | N | P | P | P | P | P | P | P | P | P | P |
| CYS | P | N | N | 0 | N | N | N | N | N | N | N | N | N | N |
| PRO | P | P | P | N | 0 | P | P | N | P | P | P | P | N | N |
| LEU | P | P | P | N | P | 0 | P | P | P | P | P | P | P | P |
| MET | P | P | P | N | P | P | 0 | P | P | P | P | N | P | P |
| TRP | P | P | P | N | N | P | P | 0 | P | P | P | N | P | P |
| PHE | P | P | P | N | P | P | P | P | 0 | P | P | P | P | P |
| SER | P | P | P | N | P | P | P | P | P | 0 | P | P | P | P |
| THR | P | P | P | N | P | P | P | P | P | P | 0 | P | P | P |
| TYR | P | P | P | N | P | P | N | N | P | P | P | 0 | N | N |
| ASN | P | P | P | N | P | P | P | P | P | P | P | N | 0 | P |
| GLN | P | P | P | N | P | P | P | P | P | P | P | N | P | 0 |

## 4.3 Applying the model to predict precipitation in cell media

To better depict the effect of addition of one amino acid on the solubility of another, a heatmap was constructed as shown in Figure 7. The amino acid on the x axis is added to the binary mixture of the amino acid in water on the y axis. Heatmap values represent deviation from the amino acid solubility in binary systems. Red was used to represent a reduction in solubility and blue was used to represent an increase in solubility in our heat map. Due to lack of interaction parameters between certain functional groups in our dataset, predictions on ternary system solubilities cannot be made for some systems. Hence, the boxes remained blank (a black box).

This heat map 'matrix' represents the effect of an amino acid on the solubility of another. This was quantified (Equation 14) by calculating the trend of solubility for an amino acid at its solubility point, when a small amount of another amino acid is added. For example, a solution of serine at its solubility point, will see an increase in solubility when some glutamine is added to the solution. This is represented by the value of +0.56. As explained in Section 4.2, this outcome arises from the interaction parameters between the functional groups of serine and glutamine that have a net effect to reduce the activity of serine in the solution, when computed via UNIFAC. The actual value reported in the matrix (here 0.56), is derived from considering the change in solubility of the amino acid at solubility when a small amount of a second amino acid is added. In our implementation, to be consistent and cognizant of the fact that amino acids all have different solubility points, 1/1000$^{th}$ of the secondary amino acid's maximum solubility in water is used to calculate the change in solubility of the primary amino acid. So, the solubility of serine is calculated, as a small amount of glutamine (1/1000$^{th}$ of its solubility in water in mol/kg) is added, and this new solubility is compared to the solubility of pure serine in water (Equation 14 below).

$$Number \in heatmap = \left(\frac{Change \in solubility\ for\ AminoAcid_A}{Small\ amount\ of AminoAcid_B\ added}\right), normalized\ for\ solubility\ of\ AminoAcid_A$$

$$Number \in heatmap = \frac{\dfrac{S_{A-new} - S_{A \in water}}{\left(\frac{1}{1000} S_{B \in water}\right)}}{S_{A \in water}} \tag{14}$$

This heat map provides a heuristic visualization for how the addition of a small amount of a second amino acid to a saturated solution of another amino in water changes that solubility. Modern fed-batch processes at both industrial and academic scales, are developed to have a glucose feed plus one or more feed media comprising chiefly of select amino acids [35,36]. While feeds are crucial for the longevity of the culture, having all amino acids in one concentrated feed creates constraints due to possible precipitation. One of the potential applications of this heatmap is to create 'pots' of amino acids that help to mutually improve each other's solubility. By mixing compatible amino acids together to create separate cell culture media feeds with enhanced solubilities of amino acids, more concentrated solutions of specific amino acid combinations could be developed and used as separate or independent feeds.

For example, by using data from the heatmap, we generated 30 pots of 3 amino acids that mutually increase each other's solubility, enabling a potentially unprecedented concentration of amino acids in the separate feeds. The groups of 3 different amino acids are listed in Table 8. This will be of particular use for amino acids such as leucine and tryptophan which have low solubilities but may undergo enhanced solubility levels by mixing with the two other amino acids in that grouping.

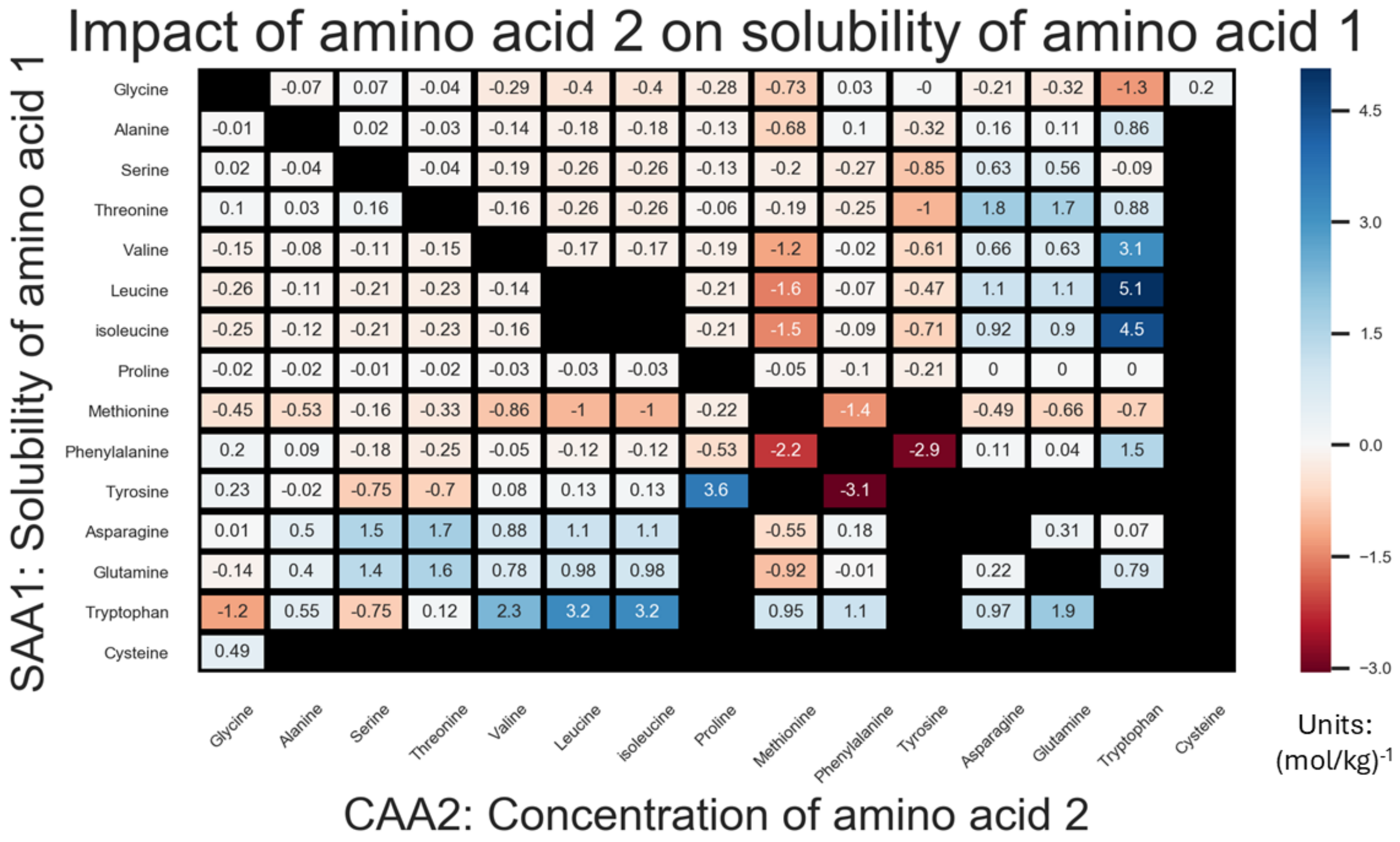

**Figure 7:** Heatmap representing the change in solubility of amino acids in ternary systems. Red represents a drop in amino acid solubility and blue represents an increase. These calculations were made at a temperature of 298 K. Equation 14 was used to calculate the impact of adding a small amount of amino acid 2 (1/1000$^{th}$ amino acid 2 solubility limit) to a saturated solution of amino acid 1. Additional ternary system data will be required to model the systems in the black cells.

**Table 8**: Model suggested amino acid clusters to improve the solubility of amino acids in quaternary systems as predicted by Figure 7.

| System number | Amino acid 1 | Amino acid 2 | Amino acid 3 |
|---|---|---|---|
| 1 | Alanine | Tryptophan | Phenylalanine |
| 2 | Alanine | Tryptophan | Asparagine |
| 3 | Alanine | Tryptophan | Glutamine |
| 4 | Alanine | Phenylalanine | Asparagine |
| 5 | Alanine | Phenylalanine | Glutamine |
| 6 | Alanine | Serine | Asparagine |
| 7 | Alanine | Serine | Glutamine |
| 8 | Alanine | Asparagine | Glutamine |
| 9 | Valine | Tryptophan | Phenylalanine |
| 10 | Valine | Tryptophan | Asparagine |
| 11 | Valine | Tryptophan | Glutamine |
| 12 | Valine | Phenylalanine | Asparagine |
| 13 | Valine | Phenylalanine | Glutamine |
| 14 | Valine | Asparagine | Glutamine |
| 15 | Leucine | Tryptophan | Phenylalanine |
| 16 | Leucine | Tryptophan | Glutamine |
| 17 | Leucine | Tryptophan | Asparagine |
| 18 | Leucine | Phenylalanine | Asparagine |
| 19 | Leucine | Phenylalanine | Glutamine |
| 20 | Leucine | Asparagine | Glutamine |
| 21 | Tryptophan | Phenylalanine | Asparagine |
| 22 | Tryptophan | Phenylalanine | Glutamine |
| 23 | Tryptophan | Threonine | Asparagine |
| 24 | Tryptophan | Threonine | Glutamine |
| 25 | Tryptophan | Asparagine | Glutamine |
| 26 | Phenylalanine | Asparagine | Glutamine |
| 27 | Serine | Threonine | Asparagine |
| 28 | Serine | Threonine | Glutamine |
| 29 | Serine | Asparagine | Glutamine |
| 30 | Threonine | Asparagine | Glutamine |

## 4.4 Multi component predictions

This study has enabled us to accurately correlate solubility data for multiple binary and ternary systems of amino acids. However, cell culture media usually contain numerous amino acids. Thus, it would be beneficial to use the model to predict amino acid combinations beyond ternary systems to assist in the design of cell culture media and feeds. This should be possible in theory as UNIFAC is a group-contribution-based method and is agnostic to the number of components used as long as interaction parameters are available. Indeed, this was one of the reasons why we adopted UNIFAC. To test whether this approach could generalize to mixtures with more than two amino acids, we looked to evaluate our model on an appropriate dataset.

An experimental data set was found involving quaternary systems of 3 amino acids [31]. Next, the interaction parameters regressed in this paper were used to make predictions on the solubility of l-leucine, l-valine, and l-alanine in water. These three amino acids were appropriate to use as all of the interactions between their functional groups had been fit to our model. The experimental data measured the solubility of one amino acid as a function of the concentration of the other two amino acids. Two points for each amino acid were measured, providing 6 datapoints from which to validate our model. The solubility of the primary amino acid was predicted and then compared to the measured value. The binary solubility of the primary amino acid was also plotted to show the difference in solubility brought about by incorporating additional amino acids.

The prediction values are shown in Figure 8, for all 6 example points along with the experimental measurements. Interestingly, the predicted values (blue bars) are in close agreement with the experimental values (orange bars) and furthermore show quite a change from the binary solubility (grey bars), in which the amino acid is alone in water. For the first example of alanine, the model was able to predict the solubility of the amino acid in the presence of leucine and valine. Moreover, the model successfully predicted a significant drop in the solubility of alanine following leucine and valine addition (Figure 8A). For the second example dataset, the model was able to predict a further decrease in solubility of valine once the amounts of alanine and leucine were added. The behavior of the other quaternary systems was captured as well, showing both a decrease in solubility compared to the binary system and an accurate prediction of the effect of changing the concentrations of 2 added amino acids.

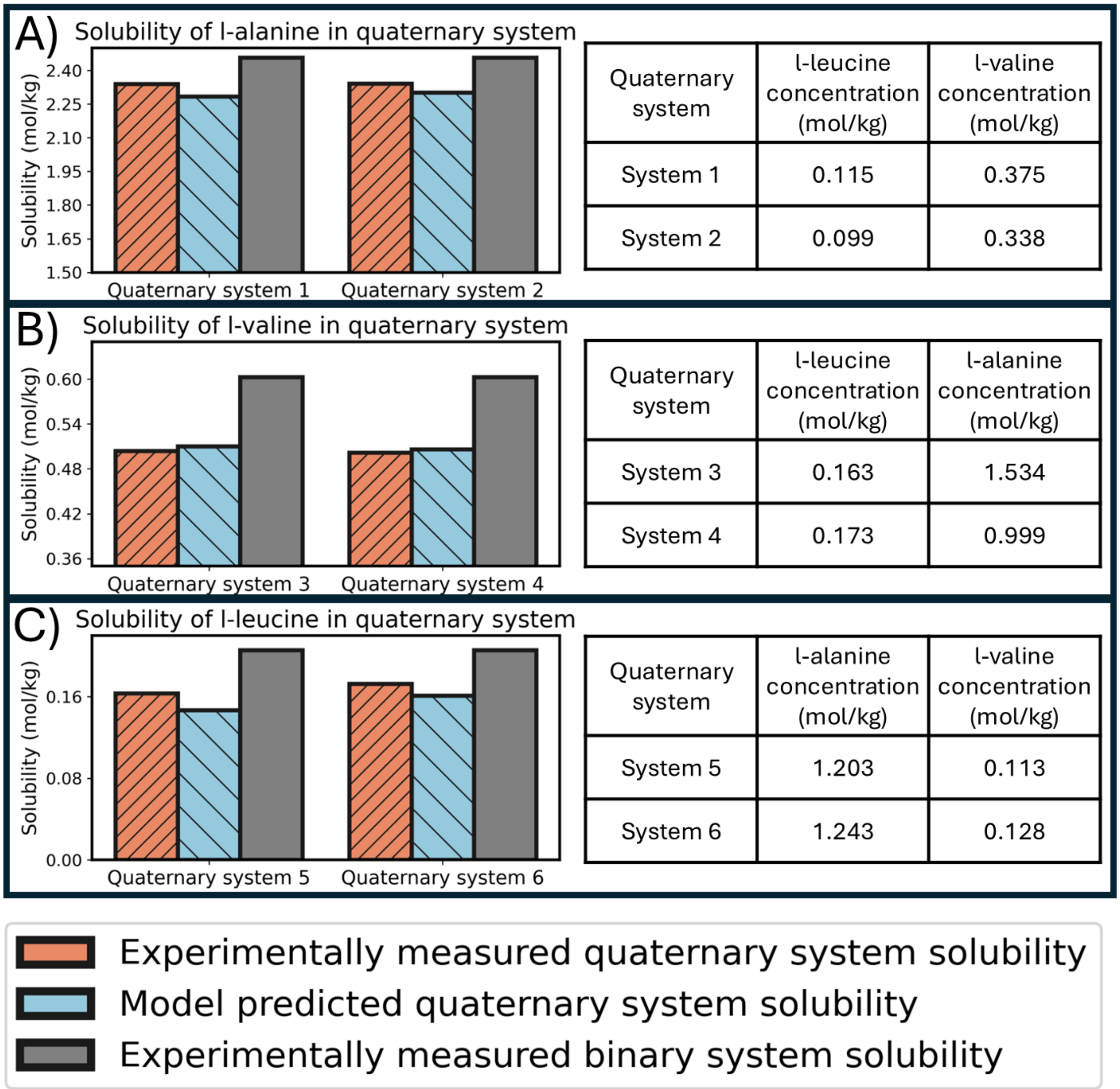

| Quaternary system | l-leucine concentration (mol/kg) | l-valine concentration (mol/kg) |
|---|---|---|
| System 1 | 0.115 | 0.375 |
| System 2 | 0.099 | 0.338 |

| Quaternary system | l-leucine concentration (mol/kg) | l-alanine concentration (mol/kg) |
|---|---|---|
| System 3 | 0.163 | 1.534 |
| System 4 | 0.173 | 0.999 |

| Quaternary system | l-alanine concentration (mol/kg) | l-valine concentration (mol/kg) |
|---|---|---|
| System 5 | 1.203 | 0.113 |
| System 6 | 1.243 | 0.128 |

**Figure 8:** Predictions of amino acid solubility in quaternary systems. Example are shown for quaternary systems consisting of alanine, leucine, valine, and water for two different concentrations of the three amino acid pairs. A) Solubility of l-alanine in systems of l-leucine, l-valine, and water. B) Solubility of l-valine in systems of l-leucine, l-alanine, and water. C) Solubility of l-leucine in systems of l-alanine, l-valine, and water. The predictions of solubility of amino acids in quaternary systems (blue bars) are very close to the experimentally measured quaternary system solubility; thus, demonstrating the predictive capabilities of the model.

These predictions are highly promising as they present evidence for our model being generalizable to even larger combinations of amino acid systems. Furthermore, the model is accurate for data sets generated in our lab as well as those that were collected from the literature.

Performing additional experiments on more ternary systems can help to further develop our heat map (Figure 7) for additional complex amino acids. Regressing new parameters on this data will help make the model apply to an even larger collection of and make it appropriate for higher order multicomponent amino acid systems. Section 4.5 delves into the future potential model modifications and directions.

## 4.5 Road to development of a digital twin for media formulation

The work presented in this manuscript represents a systematic effort to model solubility limits in amino acid systems that are critical nutrients in cell culture media. The model in its current form can aid with basal media design. Design and optimization of basal media necessitates the exploration of a large space of nutrient compositions. While automated cell culturing systems have been developed to perform high throughput experiments, the number of possible experiments is still large. This model will significantly limit the number of possible amino acid concentrations that need to be evaluated, reducing media development time and experiments that would have resulted in unwanted precipitation of target amino acids. The current form of the model can also aid with the design of feed media that often consists of very high concentrations of nutrients with even more pronounced challenges of precipitation. Separating feed media into multiple pots is commonly performed to dissolve nutrients with low solubilities such as tyrosine in a separate feed media [37]. The UNIFAC based model presented in this study may be used to implement and optimize multi-pot feed media formulations in which the most compatible set of nutrients can be dissolved together at the highest possible concentrations. To demonstrate this, the data from the heat map in Figure 7 has been used to determine clusters of amino acids that improve each other's solubility in Table 8.

Cell culture media often contains 50 to 100 compounds that are exposed to various pH and temperature conditions due to storage and operational needs. The UNIFAC model has historically been used to predict liquid phase activity coefficients at varying temperatures [38]. If sufficient solubility data at various temperatures is available for the crucial components of cell culture media, then this can be included into the model in the future. Incorporation of this data will help predict if storage at 4 °C or bioreactor operation at 37 °C will result in precipitation. Cell culture media is also exposed to a large difference in pH while it is prepared from its powered form and during operation in the bioreactors and increasing the pH in the feed media has also proved as a strategy to dissolve a few media components such as tyrosine [37]. The ability to incorporate the effect of pH in predictions of solubilities of amino acids in multicomponent systems will also prove to be useful while designing cell culture media. This has been done for binary systems of amino acids in water by modeling short -range contributions to activity coefficient via UNIFAC model with the Debye-Huckel equation to model long range contributions to the activity coefficient [39]. Activity coefficient models that can model long range interactions at high ionic concentrations have also recently been developed [40]. Future work can integrate these models with UNIFAC to evaluate the effect of pH on solubility of amino acids in ternary systems.

Cell culture media is complex and understanding solubility of components in these formulations is challenging and currently primarily done through pain-staking experimentation. However, this can be overcome by combining the efforts of the current studies with the above mentioned aspects into future modeling efforts towards digital twin for media formulation.

# 5 Conclusion

Through this study, we built a robust model to predict amino acid solubility in water for potential applications as a digital twin in the biopharmaceutical industry and beyond. This was done by regressing a functional-group based model, UNIFAC, on data that was collected in our lab and in literature. A set of UNIFAC parameters to accurately predict ternary system phase diagrams for amino acids was determined. From this, a system for predicting solubilities of amino acids in ternary system with the potential to make predictions on multicomponent systems was developed and validated using quaternary data. A heatmap showing the effect of adding one amino acid to another amino acid at its solubility limit that should be useful to those developing cell culture media formulations to aid in deciding which amino acids adversely impact the solubility of other amino acids.

The modifications that need to be incorporated to consider the effect of pH, temperature, and validating multicomponent predictions were discussed. The past decade has seen a significant push to develop various types of mathematical models to reduce, and judiciously direct experiments related to media design and other aspects of upstream process intensification in order to optimize management of resources and manpower, resources, reduce costs, and speed time to market for biopharma products [41,42]. Cell culture media process development will benefit enormously from additional in silico tools as discussed in this study. This work is an important step towards a more effective mammalian cell culture and biomanufacturing future by applying a mechanistic understanding of the thermodynamic forces involved in the components of cell culture media to design and implement in silico tools that can enhance and speed the design of media and feed components for improve mammalian cell culture biomanufacturing performance.

## Data availability statement

The data that support the findings of this study are available from the corresponding author upon reasonable request. We have also added the raw data of activity coefficients and experimental solubilities to the extended materials.

## Acknowledgments

This work was funded and supported by the Advanced Mammalian Biomanufacturing Innovation Center (AMBIC) through the Industry – University Cooperative Research Center Program of the U.S. National Science Foundation (grant numbers 1624684, 1624698, 2100075). We would like to express our gratitude to all AMBIC member companies for their mentorship and financial support.

## Author contributions

**Jayanth Venkatarama Reddy:** Conceptualization, Data Curation, Formal analysis, Investigation, Methodology, Validation, Visualization, Writing - original draft, Writing - review & editing; **Nelson Ndahiro:** Conceptualization, Data Curation, Formal analysis, Investigation, Methodology, Validation, Visualization, Writing - original draft, Writing - review & editing; **Lateef Aliyu:** Data Curation, Formal analysis, Investigation, Methodology; **Ashwin Dravid:** Conceptualization, Data Curation, Formal analysis, Investigation, Writing - review & editing **Tianxin Zhang:** Conceptualization, Data Curation, Investigation **Jinke Wu:** Data Curation, Methodology **Marc Donohue:** Conceptualization, Funding acquisition, Methodology, Project administration, Supervision, Validation, Writing - original draft, Writing - review & editing **Michael J. Betenbaugh:** Conceptualization, Funding acquisition, Project administration, Writing - original draft, Writing - review & editing.

# Supplementary section: Computational predictions of nutrient precipitation for intensified cell culture media via amino acid solution thermodynamics

# Figures

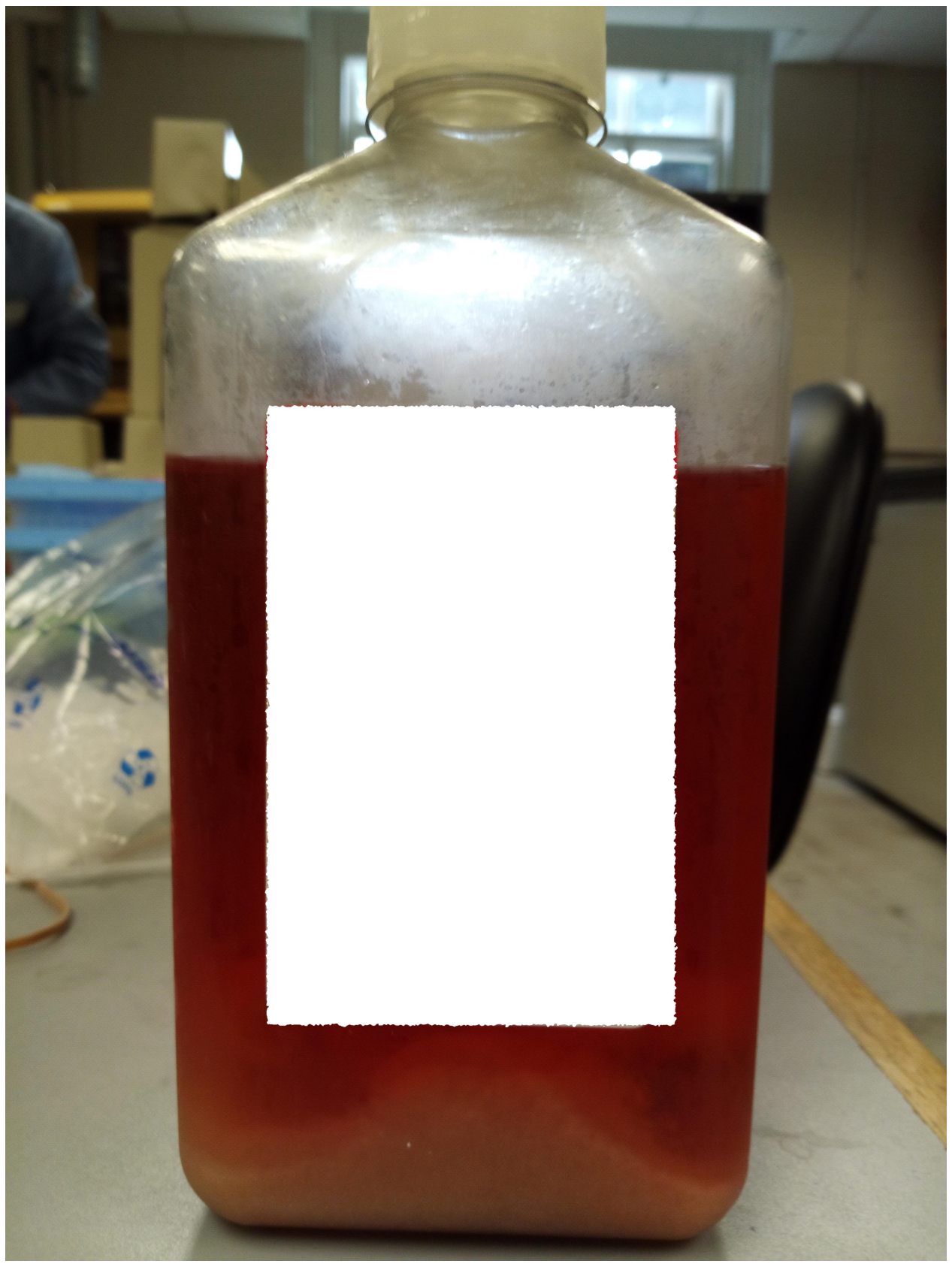

**Figure S1:** Precipitate observed in cell culture media

# Tables

**Table S1:** Isoelectric points of amino acids

| **No** | **Amino acid** | **pI** |
|---|---|---|
| 1 | Alanine | 6.11 |
| 2 | Asparagine | 5.41 |
| 3 | Cysteine | 5.02 |
| 4 | Glutamine | 5.65 |
| 5 | Glycine | 6.06 |
| 6 | Phenylalanine | 5.91 |
| 7 | isoleucine | 6.04 |
| 8 | Leucine | 6.04 |
| 9 | Methionine | 5.74 |
| 10 | Proline | 6.30 |
| 11 | Serine | 5.68 |
| 12 | Threonine | 5.6 |
| 13 | Tryptophan | 5.88 |
| 14 | Tyrosine | 5.63 |
| 15 | Valine | 6.02 |
| 16 | Histidine | 7.64 |
| 17 | Arginine | 10.76 |
| 18 | Aspartic acid | 2.87 |
| 19 | Glutamic acid | 3.08 |
| 20 | Lysine | 9.47 |

**Table S2:** Interaction parameters for functional groups determined from binary system activity coefficient data. Better representative values were determined by utilizing ternary system data as well as binary system activity system data shown in Table 5 of the manuscript. This table of interaction parameters were used to make Figure 3 and Figure 4 to demonstrate the drawbacks of using only binary system activity coefficients to make predictions of solubility in ternary systems.

| **Main group** | $H_2O$ | Alpha-CH2 | NH2 | COOH | Sc-CH3 | OH | ACOH |
|---|---|---|---|---|---|---|---|
| $H_2O$ | 0 | 71.1 | 280.6 | -82.8 | -949 | 373.4 | -102 |
| Alpha-CH2 | 1547.2 | 0 | 302.8 | -425.2 | -1022.9 | -35.6 | -218 |
| NH2 | -97.5 | 1134.4 | 0 | -679 | -698.2 | 204.4 | -723.3 |
| COOH | -378.8 | -200.8 | -314.7 | 0 | -799.0 | 15.1 | -317.1 |
| Sc-CH3 | -231.6 | 2851 | -463 | 0.776 | 0 | 2073.8 | -1065.2 |
| OH | 289.9 | -521.9 | 1459.5 | 49.4 | 1484.5 | 0 | NA |
| ACOH | -213.4 | 59.9 | -564.6 | -1010.6 | -361.3 | NA | 0 |
| ACH | -171.5 | -438.5 | -101.1 | -160.5 | 187.5 | NA | -964.6 |
| CONH2 | -99.3 | 1232.8 | 742.3 | -2.136 | -1540.8 | NA | NA |
| CH3S | -361.6 | -299.7 | 565.9 | -1358.1 | -540.3 | NA | NA |
| CH2NH | 749.35 | 1037.3 | NA | -651.7 | -1440.6 | NA | NA |
| CH2SH | 351.7 | 1769.1 | 355.3 | -504.4 | NA | NA | NA |
| Histidine ring/IMIDAZOL | -132.2 | -265.8 | -306.3 | -244.7 | -60.5 | NA | NA |
| TriN | 431.2 | -333.1 | 2216 | 173.5 | -1411.9 | NA | NA |
| ACHNH | -699 | -817.5 | -949.1 | -889.5 | -178.2 | NA | NA |
| | | | | | | | |

| **Main group** | ACH | CONH2 | CH3S | CH2NH | CH2SH | Histidine ring/IMIDAZOL | TriN |
|---|---|---|---|---|---|---|---|
| $H_2O$ | -662 | -937 | -436 | 6.4 | 404.7 | 360.4 | 34.1 |
| Alpha-CH2 | -1241 | -1249 | 566.9 | -77.9 | -376.4 | -469.2 | -215.9 |
| NH2 | -981.6 | -447.0 | 187.6 | NA | -1.09 | -110.6 | -2.89 |
| COOH | -855.3 | 444.4 | 483.1 | -633.7 | 5.47 | -160.1 | -32.33 |
| Sc-CH3 | -870.5 | -1914.4 | -168.1 | -505.3 | NA | -591.4 | -381.9 |
| OH | NA | NA | NA | NA | NA | NA | NA |
| ACOH | -67.96 | NA | NA | NA | NA | NA | NA |
| ACH | 0 | NA | NA | NA | NA | NA | NA |
| CONH2 | NA | 0 | NA | NA | NA | NA | NA |
| CH3S | NA | NA | 0 | NA | NA | NA | NA |
| CH2NH | NA | NA | NA | 0 | NA | NA | NA |
| CH2SH | NA | NA | NA | NA | 0 | NA | NA |
| Histidine ring/IMIDAZOL | NA | NA | NA | NA | NA | 0 | NA |
| TriN | NA | NA | NA | NA | NA | NA | 0 |
| ACHNH | -277.7 | NA | NA | NA | NA | NA | NA |
| | | | | | | | |
| **Main group** | ACHNH | | | | | | |
| $H_2O$ | 527.7 | | | | | | |
| Alpha-CH2 | -319.1 | | | | | | |
| NH2 | 22.57 | | | | | | |
| COOH | 23.66 | | | | | | |
| Sc-CH3 | -454.1 | | | | | | |
| OH | NA | | | | | | |
| ACOH | NA | | | | | | |
| ACH | -27.8 | | | | | | |
| CONH2 | NA | | | | | | |
| CH3S | NA | | | | | | |
| CH2NH | NA | | | | | | |
| CH2SH | NA | | | | | | |
| Histidine ring/IMIDAZOL | NA | | | | | | |
| TriN | NA | | | | | | |
| ACHNH | 0 | | | | | | |